\documentclass[twocolumn]{aastex631}

\usepackage{graphicx}	
\usepackage{amsmath}	
\usepackage{amssymb}	
\usepackage{comment}
\usepackage{booktabs}
\usepackage{upgreek}
\newcommand{\mum}{$\upmu$m }

\shorttitle{A Lack of Variability on WASP-43b}
\shortauthors{Murphy et al. 2022}

\begin{document}

\title{A Lack of Variability Between Repeated Spitzer Phase Curves of WASP-43b}

\correspondingauthor{Matthew M. Murphy}
\email{mmmurphy@arizona.edu}

\author[0000-0002-8517-8857]{Matthew M. Murphy}
\affiliation{Steward Observatory, University of Arizona, Tucson, AZ, 85705, USA}

\author[0000-0002-9539-4203]{Thomas G. Beatty}
\affiliation{Department of Astronomy, University of Wisconsin--Madison, Madison, WI, 53703, USA}

\author[0000-0001-8206-2165]{Michael T. Roman}
\affiliation{University of Leicester, School of Physics and Astronomy, University Road, Leicester, LE1 7RH, UK}
\affiliation{Facultad de Ingeniería y Ciencias, Universidad Adolfo Ibáñez, Av. Diagonal las Torres 2640, Peñalolén, Santiago, Chile}

\author[0000-0003-0217-3880]{Isaac Malsky}
\affiliation{Department of Astronomy and Astrophysics, University of Michigan, Ann Arbor, MI, 48109, USA}

\author{Alex Wingate}
\affiliation{Department of Astronomy and Astrophysics, University of Michigan, Ann Arbor, MI, 48109, USA}

\author{Grace Ochs}
\affiliation{Department of Astronomy and Astrophysics, University of Michigan, Ann Arbor, MI, 48109, USA}

\author{L. Cinque}
\affiliation{Department of Astronomy and Astrophysics, University of Michigan, Ann Arbor, MI, 48109, USA}

\author[0000-0002-6980-052X]{Hayley Beltz}
\affiliation{Department of Astronomy and Astrophysics, University of Michigan, Ann Arbor, MI, 48109, USA}

\author[0000-0003-3963-9672]{Emily Rauscher}
\affiliation{Department of Astronomy and Astrophysics, University of Michigan, Ann Arbor, MI, 48109, USA}

\author[0000-0002-1337-9051]{Eliza M.-R. Kempton}
\affiliation{Department of Astronomy, University of Maryland, College Park, MD, 20742, USA}

\author[0000-0002-7352-7941]{Kevin B. Stevenson}
\affiliation{Johns Hopkins APL, Laurel, MD, 20723, USA}

\begin{abstract}
Though the global atmospheres of hot Jupiters have been extensively studied using phase curve observations, the level of time variability in these data is not well constrained. To investigate possible time variability in a planetary phase curve, we observed two full-orbit phase curves of the hot Jupiter WASP-43b at 4.5 \mum using the \textit{Spitzer Space Telescope}, and reanalyzed a previous 4.5 \mum phase curve from \cite{Stev16}. We find no significant time variability between these three phase curves, which span timescales of weeks to years. The three observations are best fit by a single phase curve with an eclipse depth of 3907 $\pm$ 85 ppm, a dayside-integrated brightness temperature of 1479 $\pm$ 13 K, a nightside-integrated brightness temperature of 755 $\pm$ 46 K, and an eastward-shifted peak of 10.4$^\circ$ $\pm$ 1.8$^\circ$ degrees. To model our observations, we performed 3D general circulation model simulations of WASP-43b with simple cloud models of various vertical extents. In comparing these simulations to our observations, we find that WASP-43b likely has a cloudy nightside that transitions to a relatively cloud-free dayside. We estimate that any change in WASP-43b's vertical cloud thickness of more than three pressure scale heights is inconsistent with our observed upper limit on variation. These observations, therefore, indicate that WASP-43b's clouds are stable in their vertical and spatial extent over timescales up to several years. These results strongly suggest that atmospheric properties derived from previous, single \textit{Spitzer} phase curve observations of hot Jupiters likely show us the equilibrium properties of these atmospheres.
\end{abstract}

\section{Introduction}
\label{sec:intro}

Weather and climate are intrinsically linked to interpretations of planetary properties deduced from remote observations. The history of observations within our own Solar System tells cautionary tales for neglecting weather phenomena. For example, early observers of Mars noticed seasonal variations in dark spots on Mars' surface, and interpreted them as the coming and going of vast patches of vegetation \citep[see, e.g.,][]{sinton1957, sinton1959}. This variation was later revealed to just be the seasonal obscuration and uncovering of the rocky, inanimate surface by dust storms. Exoplanetary science is entering a revolutionary era for high-precision spectroscopy of exoplanet atmospheres, with the potential for similar difficulties in interpreting newfound phenomena. 

Our understanding of an exoplanet's atmosphere is unavoidably linked to the effects of weather and climate in that atmosphere. This is because variability in atmospheric mixing, cloud formation, and circulation can drive the planet's observable properties to change in time. As a result, revealing the true nature of the atmosphere may require repeated observations. In an atmosphere that varies in time, properties derived from only a single observation provide only a snapshot state that does not necessarily represent the general nature of a planet.

Some early simulations of circulation in hot Jupiter atmospheres predicted that atmospheric dynamics would naturally produce features like polar vortices, zonal jets, and an asymmetric temperature field. These weather features were predicted to cause variations in a planet's emission properties, depending on the planet's wind speeds and irradiation environment \citep{cho2003, cho2008}. \cite{rauscher2008} simulated how these weather features, particularly polar vortices, could cause a planet's phase curve to change over time. They predicted that weather would cause the phase curve to exhibit amplitude modulations upwards of 25\% over the course of consecutive orbits. 

Other theoretical work, however, has not predicted strong variations driven by the planet's climate \citep{showman2002, cooper2005, dobbsdixon2008}. Some models further predicted only minor variations in observable quantities over time. For example, \cite{showman2009}'s 3D, radiative-dynamical models of the hot Jupiters HD 189733b and HD 209458b exhibited only $\sim$1\% variability in the eclipse depth as measured in all \textit{Spitzer} channels. \cite{dobbsdixon2013}'s simulations of HD 189733b exhibited very stable dayside emission, varying by less than 0.1\% over 30 orbits for all \textit{Spitzer} channels. Similarly, \cite{menou2020}'s high-resolution models of HD 209458b exhibited less than 2\% variability in the infrared dayside-integrated flux. Finally, \cite{komacek2020}'s simulations of phase curve observations of HD 209458b predicted upper limits on eclipse depth variability of 2\% and on phase curve amplitude of 1\%. 

It is worth noting that all of the above models were cloud-free. Adding clouds to a planet's atmosphere is known to change its emission characteristics significantly compared to a cloud-free atmosphere \citep[e.g.][]{Mendonca2018a, venot2020, roman2021clouds}, including inducing variability in the observed emission \citep[e.g.][]{parmentier2013, line2016, lines2018, powell2018, jackson2019}. 
In addition to clouds, magnetic interactions between a planet's magnetic field and ions embedded in a planet's atmospheric winds may also cause variability. This would be true for hot Jupiters warm enough that they experience significant thermal ionization on their daysides \citep{rogers2017, hindle2019, hindle2021}. 

Early infrared observations of hot Jupiter atmospheres agreed with the lack of variability predicted by cloud-free modeling. \cite{agol2010}'s repeated secondary eclipse observations of HD 189733b using \textit{Spitzer}/InfraRed Array Camera (IRAC) at 8~\mum placed an upper limit of 2.7\% on the variability of HD 189733b's eclipse depth at that wavelength. \cite{crossfield2012}'s repeated transit and eclipse observations of HD 209458b using \textit{Spitzer}/MIPS at 24~\mum found no significant variations in either the transit or eclipse depth. \cite{kilpatrick2020} found no variations in the eclipse depths of HD 189733b or HD 209458b at both 3.6~\mum and 4.5~\mum using \textit{Spitzer}/IRAC. Similarly, analyses by \cite{wong2014xo3}, \cite{ingalls2016xo3}, and \cite{bell2021} find no significant evidence for variability in XO-3b's eclipse depth at 4.5~\mum.

On the other hand, optical observations of hot Jupiters have seen significant variability. The \textit{Kepler} phase curve of HAT-P-7b exhibits large variations in its shape and offset over tens to hundreds of days \citep{armstrong2016}. Similarly, the \textit{Kepler} phase curve of Kepler-76b shows variation in its eclipse depth of $\sim$40\% on timescales of tens of days \citep{jackson2019}. \cite{bell2019} and \cite{vonessen2019} also find variation in WASP-12b's eclipse depth and phase curve offset, but the authors suggest that these changes may be the result of WASP-12b transferring mass to its star, rather than intrinsic atmospheric variations. Long-term TESS observations of WASP-12b have not found significant variability in the planet's eclipse depth over a timescale of couple of years \citep{Wong2022}. 

It is unclear whether the observed variation at optical wavelengths is actually due to weather in the planet's atmosphere. Optical observations are more vulnerable to stellar contamination than infrared observations, which can introduce time variability into long-duration time series. For example, \cite{lally2022} showed that the observed variability in HAT-P-7b's phase curve can also be explained by supergranulation on the host star.

To date, there have been no published observing programs designed to search for phase curve variability at medium infrared wavelengths, such as those probed by \textit{Spitzer}/IRAC. Repeated phase curve observations have been taken using \textit{Spitzer} to help correct the strong systematics that are often present in 3.6~\mum data (e.g., WASP-12b \citep{bell2019}, WASP-76b \citep{may2021wasp76b}, and WASP-52b \citep{May2022}) and to measure low-amplitude phase curve signals (e.g., K2-141b \citep{zieba2022k2141b}), but it is difficult to probe variability with these data sets. Repeat phase curve observations at medium infrared wavelengths present a bridge between existing constraints on variability, and a new test on the variable nature of hot Jupiter atmospheres. We set out to do this by observing two new phase curves of the hot Jupiter WASP-43b using \textit{Spitzer}/IRAC at 4.5~\mum. Together with a previous phase curve from \cite{Stev16}, these observations provide three epochs over which to sample WASP-43b's global weather. 

WASP-43b is a hot Jupiter orbiting the bright (K=9.2) K7 star WASP-43, discovered in 2011 by \cite{Hellier11}. The planet is on a short 19.5 hour orbital period \citep{Hellier11} which drives a dayside brightness temperature of 1485 $\pm$ 41 K \citep{May2020}. Previous observations of WASP-43b suggest it has a cloudy nightside \citep{Stev16, kataria2015}, and models further predict WASP-43b should have clouds on its dayside as well \citep{Helling20}. This expected cloudiness, in addition to the planet's short period and high signal-to-noise ratio in phase curve observations \citep{Stev16}, make WASP-43b an ideal target for our investigation into potential infrared variability.

\section{Observations \& Data Reduction}
\label{sec:obs}

We obtained two full-orbit photometric phase curves of WASP-43b at 4.5 \mum using \textit{Spitzer}/IRAC \citep{Fazio04} separated by two weeks, on 2019 September 11 and 2019 September 26 (PID 14297). Hereafter, we will refer to these visits as ``2019a" and ``2019b", respectively. Each visit used the subarray mode with 2.0 second exposures and lasted 25.2 hours. We employed a 30-minute pre-observation before each visit's science observations to allow any detector persistence to build up to a steady state, and excluded this time from our data analysis. The science visits were split into two consecutive 12.5 hour astronomical observing requests (AORs), so that \textit{Spitzer} would repoint and position WASP-43 onto the IRAC photometric ``sweet spot" at the start of each AOR. This re-pointing process added a gap of 6 minutes between each AOR. To stabilize the pointing in each AOR, we used the PCRS peak-up mode targeted on WASP-43 itself. 

\begin{table*} \centering
\caption{Observation details}
\begin{tabular}{lcccccccl}\toprule
Label & Observation Date & Exposure time {[}s{]} & Aperture Size [pix] & Spitzer Pipeline & Total AORs & Publication \\ 
\midrule 
2014 & 2014 August 27-28 & 2.0 & 2.4  & S19.1.0 & 3 & \cite{Stev16} \\
2019a & 2019 September 11-12 & 2.0 & 2.1 & S19.2.0 & 2 & This work \\
2019b & 2019 September 26-27 & 2.0 & 2.4 & S19.2.0 & 2 & This work \\\bottomrule 
\end{tabular}
\label{tab:obsinfo}
\end{table*}

In addition to our observations, we reanalyzed the Spitzer/IRAC 4.5 \mum full-orbit phase curve curve data taken as part of observing program PID 10169 on 2014 August 27, described by \cite{Stev16}. Hereafter, we will refer to this visit as ``2014". This visit was split into three consecutive AORs. Table \ref{tab:obsinfo} provides specific details of each observing visit. 

Our data reduction process follows that of \cite{Beatty2018}. We started with the basic calibrated data (BCD) images from the Spitzer pipeline. First, we performed background subtraction by masking out the middle of each image, 10 pixels in from each edge, to remove star light, as well as the bottom row of each image since it systematically contains several \texttt{nan} values, and then computed the median value of the unmasked pixels. We then subtracted this median value from each pixel in the image. Then, we corrected for bad pixels by performing three iterations of 5-$\sigma$ clipping on each pixel's time series and replacing the values of identified bad pixels with the median value of its time series. Using these background-subtracted and bad-pixel-corrected images, we then determined the position of WASP-43 in each image using the Howell centroiding technique \citep{Howell06}. 

\begin{figure}
    \centering
    \includegraphics[width=\linewidth]{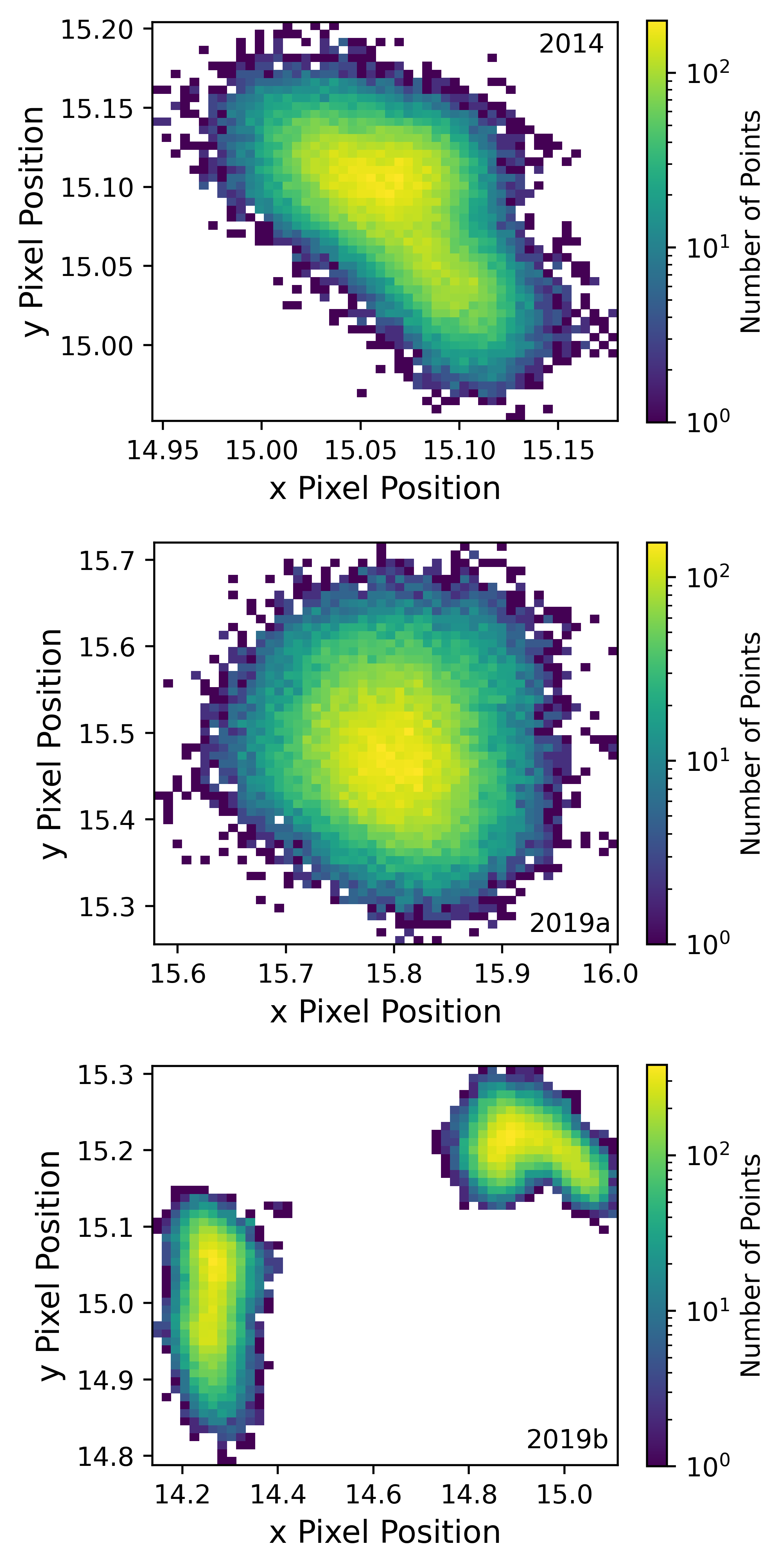}
    \caption{Pointing histograms for the 2014 visit (top), 2019a visit (middle), and 2019b visit (bottom). The 2014 visit shows considerable overlap, but there is a notable trend in centroids toward the lower right likely due to re-pointing inconsistency between its three AORs. The 2019a visit has full overlap between its two AORs. The 2019b visit has two distinct centroid regions. The top right region consists frames from the first AOR and the bottom left region consists of frames from the second AOR. This suggests the spacecraft was unable to return to its original pointing when starting the second AOR.  }
    \label{fig:PointingHistos}
\end{figure}

We used the background-subtracted images, prior to bad-pixel correction, for the photometric extraction. We extracted the flux within a circular aperture centred on WASP-43's position in each image. Figure~\ref{fig:PointingHistos} shows histograms of WASP-43b's position on the detector for each visit. The aperture size used depended on the data set. We chose the optimal aperture for each visit by sampling an astrophysical model on the data, which is described in Section \ref{subsec:model}, identifying which aperture size gave both the largest Bayesian likelihood and smallest data-model residual.
The 2014 and 2019b visits both preferred an aperture radius of 2.4 pixels, while the 2019a visit preferred a radius of 2.1 pixels. For reference, the median FWHMs were 0.50 pixels, 0.75 pixels, and 0.51 pixels for the 2014, 2019a, and 2019b visits, respectively. 
We then identified and removed any outlier flux measurements in each visit's normalized flux time series by clipping any flux value more than 5-$\sigma$ from the best-fit model found during aperture optimization. The 2019b light curve exhibited a nonlinear decreasing trend which, in lieu of fitting for systematics at this stage, we fitted with a third-order polynomial in this outlier-cutting model. This removed 59 of 44928 exposures from the 2014 data, 57 of 44928 exposures from the 2019a data, and 132 of 44928 exposures from the 2019b data. 

\textit{Spitzer} time series observations are susceptible to detector persistence, which causes a notable ramp effect at the start of each AOR. Each of our visits showed this initial ramp effect. This effect was identifiable by small increases in the background flux level over the first 5-10 minutes of each AOR, after which the background level reached a constant saturation level throughout the entire visit. The exact duration of this initial persistence ramp was slightly different for each AOR and each visit. We determined the end of each ramp period as the time when the background flux level appeared to flatten off at a constant median value, then cut out all data points prior to this time. This process amounted to cutting
150 points, 174 points, and 73 points from the three AORS of the 2014 visit, 250 points and 206 points from the two AORS of the 2019a visit, and 
200 points and 186 points from the two AORs of the 2019b visit. 
After these two clipping routines, we were left with 44719 points in the 2014 visit, 44771 points in the 2019a visit, and 44696 points in the 2019b visit. 

We determined the time of each exposure by assuming that each 64 frame image stack began at the \texttt{MJD\_OBS} header time and the image frames were evenly spaced between the \texttt{AINTBEG} and \texttt{AINTEND} header times. We then converted these spacecraft times to Barycentric Julian Dates in the Barycentric Dynamical Time standard (BJD\_TDB).

\section{Light Curve Model Selection}
\label{sec:fitting}

We modeled the observed light curve as a combination of a time-varying astrophysical signal and systematic effects introduced by the instrument. We denote the astrophysical signal as $A(t)$. The systematic error is a function of both time and spatial position (x,y) on the detector, and we denote it as $S(x,y,t)$. The full light curve model, which we call F$_{\text{obs}}$(t), was then
\begin{eqnarray}
F_{obs} = A (t) ~ S(x,y,t).
\end{eqnarray}

\subsection{Spitzer Systematics}
\label{subsec:spitzersystematics}

We found that our data displayed the usual systematic effects seen in \textit{Spitzer} time series. We divided these into intrapixel sensitive variations B(x,y), a linear trend in time $f_t (t)$, and linear trends in the x- and y-FWHMs, $f_{\text{x, FWHM}}$ and $f_{\text{y, FWHM}}$, respectively. All together, the model for systematics is
\begin{eqnarray}
S(x,y,t) = B(x,y) ~f_{t}(t)~  f_{\text{x, FWHM}} ~ f_{\text{y, FWHM}}.
\end{eqnarray}

Intrapixel sensitivity variations, which cause variations in flux at the subpixel scale, are the dominant source of systematic error in \textit{Spitzer's} 4.5 \mum channel. We corrected for these sensitivity variations by using Bilinearly-Interpolated Subpixel Sensitivity (BLISS) mapping \citep{Stev12} to fit for the relative sensitivity $B(x,y)$ of the detector as a function of position. \cite{May2020} have suggested that a fixed BLISS map, determined from multiple observations, can return the most precise fit, and have published a fixed sensitivity map. However, this map does not cover the full extent of our observed stellar positions. We therefore fit a BLISS map simultaneously with the other model parameters.

In addition to BLISS mapping, previous analyses have shown correlations between the measured flux and the FWHM of the stellar point response function (PRF) \citep[e.g.][]{Mendonca2018a}. 
To measure the FWHM in each frame, we used the Photometry for Orbits, Eclipses, and Transits (POET) pipeline \citep{Stev12, Cubillos2013_poet, Stev16} which fits the stellar point spread function of each image with a Gaussian function. We used these measured FWHMs to fit a linear detrending function that depended on the x- and y-width of the star 
\begin{eqnarray}
    f_{FWHM} = s_{\text{FWHM}} \left(w - w_{\text{median}} \right) + 1.0,
    \label{eqn:fwhmramp}
\end{eqnarray}
where $s_{FWHM}$ is the slope and $w$ is either the x- or y-FWHM. We used separate slopes for the x- and y-width models and left them as free parameters in the sampling.

Lastly, we fit a visit-long linear trend in time of
\begin{eqnarray}
    f_t = s_{t} \left( t - t_{\text{median}} \right) + 1.0,
    \label{eqn:timeramp}
\end{eqnarray}
where $s_t$ is the slope, which we left as a free parameter in our sampling, and $t_{\text{median}}$ is the median time of the visit. We tested this visit-long detrending against one where each AOR was given its temporal trend, and found that the visit-long detrending was preferred in terms of the Bayesian Information Criterion (BIC). Similarly, we tested several other systematic correction methods, including not using FWHM detrending $f_{FWHM}$, not using flux detrending $f_t$, using a quadratic form of $f_t$, and detrending using noise pixel values, and found that our methods described here were preferred in terms of the BIC.

\subsection{Astrophysical Model}
\label{subsec:model}

We modeled the astrophysical signal from the WASP-43 system as the combination of the star's signal ($F_\star$), which we treat as just a transit model, and the planet's signal ($F_{\text{planet}}$), which is a coupled eclipse and phase curve model. Altogether, our astrophysical model was then
\begin{eqnarray}
A(\theta, t) = F_\star (\theta, t) + F_{\text{planet}} (\theta, t).
\end{eqnarray}

Here, $\theta$ represents the astrophysical parameters for which we fit: the transit center time, orbital period, ratio of the semi-major axis to the stellar radius, inclination, ratio of the planetary and stellar radius, and all phase curve parameters described below. We did not fit for WASP-43b's eccentricity or longitude of periastron. Previous works have shown that WASP-43b's eccentricity is consistent with zero \citep{Hellier11, Gillon12, Wang13, Blecic14}, so we fixed the eccentricity to zero and the longitude of periastron to $\omega = \pi/2$. 

We modeled the planetary transit $F_\star (\theta, t)$ using the \texttt{BATMAN} python package \citep{Kreidberg15}. We used a quadratic law for stellar limb darkening and fixed the coefficients to $u_1 = 0.0796$ and $u_2 = 0.1621$. We estimated these limb darkening coefficients by matching WASP-43's stellar properties in the table of modelled limb darkening coefficients by \cite{Claret2011}, using those computed via the least-squares method.

We did not consider any variability on part of the star in our model, which is justified by a lack of stellar variability seen by previous observations of the system as well as the fact that we do not end up finding variability between our observations. 

To model the relative flux from the planet $F_{\text{planet}} (\theta, t)$, we used a combined geometric eclipse and phase curve model
\begin{eqnarray}
F_{\text{planet}} (\theta, t) = f_{\text{eclipse}} ~ f_{\text{phase curve}}.
\end{eqnarray}
The geometric eclipse model, $f_{\text{eclipse}}$, models the relative change in flux during the planet's occultation. We normalize it to be equal to one while out of eclipse and have a depth of unity, so that it physically represents the fraction of the planet's disk which is visible as a function of time. Rather than having to fit for the eclipse depth itself, this method lets the eclipse depth be naturally set by the phase curve parameters. Also, when fitting the eclipse time, we account for WASP-43b's Roemer delay of 15.1 seconds.

Our phase curve model represents the emission from the planet as a function of its orbital phase. We use a double harmonic sinusoidal model of the form
\begin{eqnarray} 
  f_{\text{phase curve}} =   f_0 + c_1 \cos \left( x_1 \right) + c_3 \cos \left( x_2 \right), 
    \label{eqn:sinepc}
\end{eqnarray}
where
\begin{eqnarray}
x_1 &=& \frac{2 \pi}{P} \left( t - t_c \right) + \pi + c_2 \frac{\pi}{180}, \\
x_2 &=& \frac{4 \pi}{P} \left( t - t_c \right) + \pi + c_4 \frac{\pi}{180}.
\end{eqnarray}
Here, $f_0$ is the vertical phase offset, $P$ is the orbital period, $t_c$ is the transit midpoint time, $c_1$ and $c_3$ are the phase amplitudes, and $c_2$ and $c_4$ are the phase offsets in units of degrees. We add a factor of $\pi$ in the argument of each cosine term to center the phase curve maximum at the secondary eclipse midpoint for an offset of zero degrees. \cite{Stev16} used a similar formula for their phase curve model and found a nonzero second harmonic in their fit of the 2014 phase curve. We later test whether our fits also show a significant second harmonic. 
 
\subsection{Fitting Scenarios}

\begin{table*}
\caption{Parameters and corresponding values used as priors in our MCMC sampling.}
\centering
\label{tab:priors}
\begin{tabular}{lccccl}\toprule  
Parameter & Units & Prior Value & Source \\\midrule 
$t_c$ & Transit Center Time [BJD$_{\text{TDB}}$] & 2456612.416074 $\pm$ 0.000044 & \cite{Wong2022}  \\
P & Orbital Period [day] & 0.813474061 $\pm$ 0.000000046 & \cite{Wong2022}\\
 a / R$_\star$ & Semi-major axis [stellar radii] & 4.867 $\pm$ 0.023 & \cite{Hoyer2016} \\
cos $\left(  i \right)$  & Inclination  & 0.137 $\pm$ 0.001 & \cite{Chen2014} \\\bottomrule 
\end{tabular}
\end{table*}

Previous observations have ruled out any decay of WASP-43b's orbit \citep{Blecic14, Murgas2014, Hoyer2016, Patra2020orbitdecay, garai2021orbitdecay}, which would naturally lead to changes in the shape of WASP-43b's phase curve over time. In addition, previous observations of this system at shorter wavelengths have not seen significant stellar activity, so we do not expect stellar contamination to be a major concern \citep{Hellier11, Gillon12, czesla2013, Murgas2014}. Therefore, we expect any variability between the phase curves to be the result of intrinsic atmospheric variability. If the planet's climate or weather is causing significant changes to the hemisphere-integrated emission at each phase over time, we should see this as changes in the shape and amplitude of the phase curve. This reasoning motivated us to use three fitting scenarios that differ in which fitting parameters are allowed to vary between visits.

In the first scenario, which we refer to as the ``individually fit" case, we fit each visit's light curve individually. We gave each light curve its own set of planetary, orbital, and phase curve parameters, and its own BLISS map. In the second scenario, which we refer to as the ``semi-shared fit" case, we simultaneously fit a shared set of planetary and orbital parameters, and BLISS map, but gave each light curve its own phase curve parameters. In the third scenario, which we refer to as the ``fully shared fit" case, we fit all parameters simultaneously so that each light curve used a single, shared set of planetary, orbital, and phase curve parameters, and a shared BLISS map. In all scenarios, we gave each light curve its own set of systematic parameters for $f_t$, $f_{x, FWHM}$, and $f_{y, FWHM}$ (see Eqns. \ref{eqn:fwhmramp} and \ref{eqn:timeramp}). As we will show in Section~\ref{sec:results}, we found no significant variation between visits so this third, ``fully shared fit" scenario is preferred. The other two fitting scenarios then served as checks that our individual data sets did not have any uncorrected systematic effects present. 

To fit the data in all three scenarios, we employed Markov Chain Monte Carlo (MCMC) sampling using the \texttt{emcee} python package \citep{Mackey2013}. We enforced Gaussian priors on only four parameters: the transit center time, orbital period, semi-major axis, and inclination. Within the MCMC sampling, we parametrized the orbital period as $log_{10} \left(P \right)$, the semi-major axis as $log_{10} \left( a / R_\star \right)$, and the inclination as $\cos \left( i \right)$, following \cite{eastman2013}. We give these prior values in Table \ref{tab:priors}. We did not enforce a radius prior since the only radius measurement at this wavelength was by \cite{Stev16}, whose data we are reanalyzing in this work. We sampled each scenario for 20,000 steps, which was over 20 times their corresponding autocorrelation times as estimated by \texttt{emcee}, using a number of walkers that was twice the number of free parameters in each case. The 2014 and 2019b individual fits needed a 1,000 step burn-in, the 2019a individual fit needed a 1,500 step burn-in, and the semi-shared and fully shared fits each needed a 5,000 step burn-in.

\section{Results}
\label{sec:results}

We found no significant variation between the three observed phase curves. The fully shared fitting case was preferred of our three fitting scenarios, meaning the three visits are best described by a single, shared phase curve rather than three different phase curves. 

\subsection{Fully Shared Fits}
\label{subsec:fullyshared}

\begin{figure*}
    \centering
    \includegraphics[width=\textwidth]{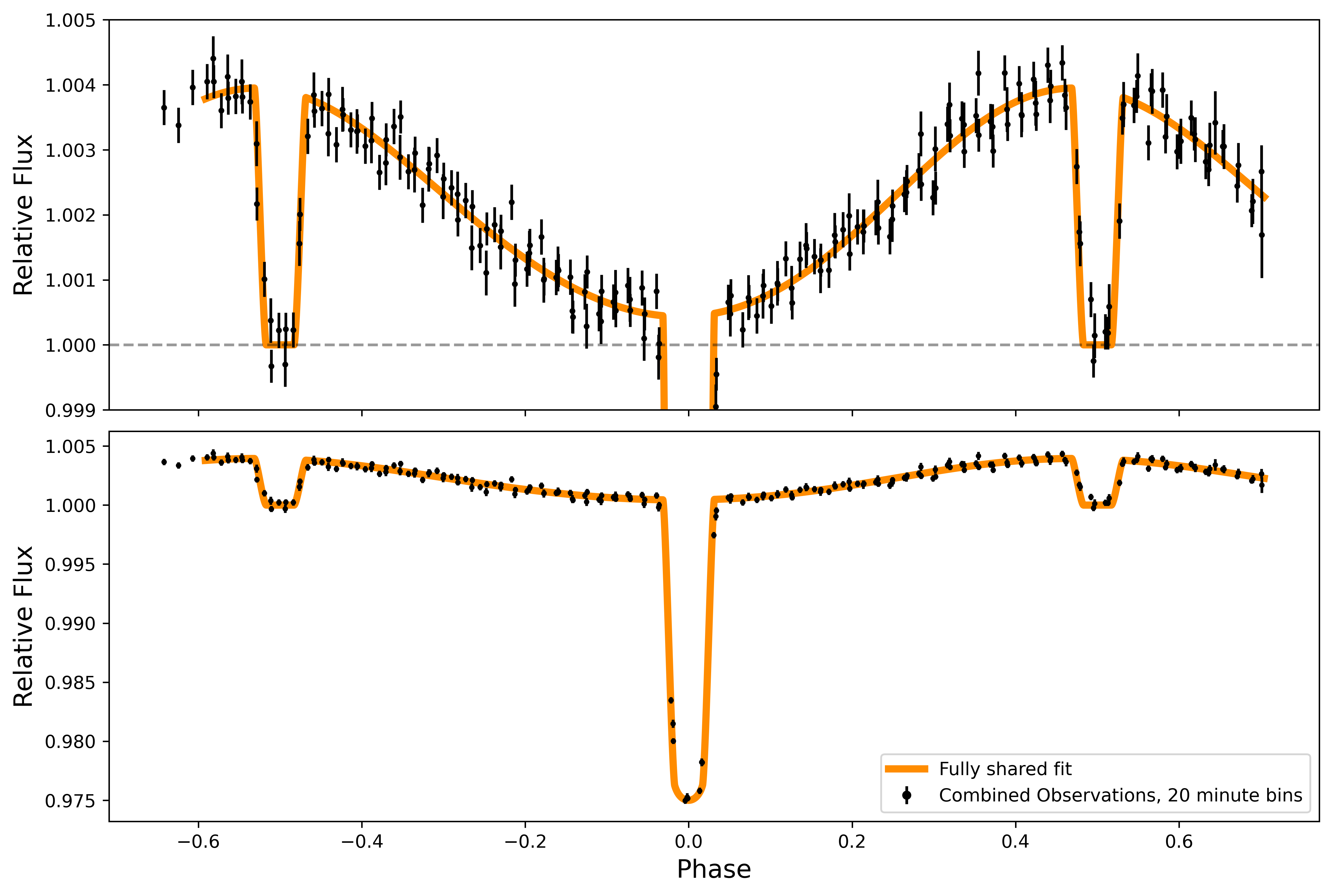}
    \caption{Emission phase curve of WASP-43b at 4.5 \mum, which is the fully shared fit in our analysis. The black data points are a combination of the data from the three visits analysed in this work, binned in 10 minute intervals. This binning is only done for presentation purposes, and only unbinned data were used during the analysis. The bottom panel shows the full light curve, while the top panel highlights the phase curve.}
    \label{fig:sys_sharedfitalldata}
\end{figure*}

As mentioned, in the fully shared fit case, we fit the three visits simultaneously with a single, shared set of planetary, orbital, and phase curve parameters, and a single BLISS map. The best-fit parameter values and relevant derived properties are given in Table~\ref{tab:fullysharedfitvals}. The full light curve is plotted in Figure~\ref{fig:sys_sharedfitalldata}, and just the phase curve component is plotted as the orange dotted line in the bottom panel of Figure~\ref{fig:pccom_combined}. Also, the raw data and FWHMs from each visit, along with each fitted model, is plotted in Figure~\ref{fig:rawplots} of the Appendix.

The fully-shared fit phase curve exhibits a small eastward hotspot offset of 10.4$^\circ$ $\pm$ 1.7$^\circ$, which is close to the 12.3$^\circ$ $\pm$ 1.0$^\circ$ eastward offset of WASP-43b's \textit{Hubble Space Telescope (HST)}/WFC3 phase curve \citep{stevenson2014}. We find a maximum phase curve value of 3936 $\pm$ 84 ppm, a nonzero minimum phase curve value of 425 $\pm$ 105 ppm, a peak-to-peak amplitude of 3511 $\pm$ 89 ppm, and an eclipse depth of 3907 $\pm$ 85 ppm. 
Using a BT-Settl model for the star \citep{BTsettl_1, BTsettl_2, BTsettl_3, BTsettl_4, BTsettl_5, BTsettl_6, BTsettl_7} and a blackbody model for the planet, we convert these phase curve extrema into integrated brightness temperatures of 1479 $\pm$ 13 K on the dayside and 755 $\pm$ 46 K on the nightside. The phase curve's second harmonic term does not strongly contribute, as $c_3$ is close to zero and $c_4$ has over 50\% uncertainty and is not well constrained. 

\begin{table*}
\caption{Best-fit Values When Fitting the Three Data Sets Simultaneously with Shared Planetary, Orbital, and Phase Curve Parameters with Only Individual Systematic Parameters. } 
\centering
\label{tab:fullysharedfitvals}
\hspace*{-2.5cm}\begin{tabular}{lcccccl}\toprule 
Parameter & Units & 2014 & 2019a & 2019b \\\midrule 
Fitting Parameters & & & &  \\
$t_c$ & Previous transit time [BJD$_{\text{TDB}}$] & 2456612.41608 $\pm$ 4 $\times$ $10^{-5}$ & $\leftarrow$ & $\leftarrow$ \\
$\log_{10} \left(P\right)$ & log Orbital period [days] & -0.08965628 $\pm$ 1 $\times$ $10^{-8}$ & $\leftarrow$ & $\leftarrow$ \\
$\log_{10} \left(a / R_\star \right)$ & log scaled semi-major axis & 0.687 $\pm$ 0.001 & $\leftarrow$ & $\leftarrow$  \\
$\cos ~i$  & Cosine of inclination angle & 0.1368 $\pm$ 0.0008 & $\leftarrow$ & $\leftarrow$  \\
$R_p / R_\star$ & Radius ratio & 0.1578 $\pm$ 0.0004 & $\leftarrow$ & $\leftarrow$ \\
$c_1$ & First harmonic amplitude & 0.00175 $\pm$ 0.00004 & $\leftarrow$ & $\leftarrow$ \\
$c_2$ & First harmonic offset [deg] & 7.5 $\pm$ 1.4 & $\leftarrow$ & $\leftarrow$ \\
$c_3$ & Second harmonic amplitude &  -1.0 $\times$ $10^{-4} \pm$ 4 $\times$ $10^{-5}$ & $\leftarrow$ & $\leftarrow$ \\
$c_4$ & Second harmonic offset [deg] & 43.2 $\pm$ 22.3 & $\leftarrow$ & $\leftarrow$ \\
$f_0$ & Phase curve vertical offset & 0.00210 $\pm$ 0.00008 & $\leftarrow$ & $\leftarrow$ \\
$s_t$ & Linear ramp slope & -0.0009 $\pm$ 0.0001 & -0.0005 $\pm$ 0.0001 & -0.0042 $\pm$ 0.0005 \\
$s_x$ & x-FWHM ramp slope & -0.025 $\pm$ 0.008 & -0.023 $\pm$ 0.004 & 0.001 $\pm$ 0.002 \\
$s_y$ & y-FWHM ramp slope & 0.024 $\pm$ 0.008 & 0.021 $\pm$ 0.005 & 0.001 $\pm$ 0.004 \\
Derived Parameters & & & & \\
$t_c$ & Current transit time [BJD$_{\text{TDB}}$] & 2456897.13201 $\pm$ 0.00004 & 2458738.83731 $\pm$ 0.00008  & 2458753.47985 $\pm$ 0.00008  \\
$t_s$ & Current eclipse time [BJD$_{\text{TDB}}$] & 2456897.53892 $\pm$ 0.00004 & 2458739.24422 $\pm$ 0.00008 & 2458753.88676 $\pm$ 0.00008 \\
P & Orbital Period [days] & 0.81347407 $\pm$ 2 $\times$ $10^{-8}$ & $\leftarrow$ & $\leftarrow$ \\
$a / R_\star$ & Scaled semi-major axis & 4.86 $\pm$ 0.01 & $\leftarrow$ & $\leftarrow$ \\
$i$ & Inclination angle [deg] & 82.133 $\pm$ 0.001 & $\leftarrow$ & $\leftarrow$ \\
$F_{\text{min}}$ & Minimum phase curve flux [ppm] & 425 $\pm$ 105 & $\leftarrow$ & $\leftarrow$ \\
$\delta_{\text{eclipse}}$ & Secondary eclipse depth [ppm] & 3907 $\pm$ 85 & $\leftarrow$ & $\leftarrow$ \\
$\Phi$ & Hotspot offset [degrees] & -10.4 $\pm$ 1.7 & $\leftarrow$ & $\leftarrow$ \\ 
$A_{P2P}$ & Peak-to-peak amplitude [ppm] & 3511 $\pm$ 89 & $\leftarrow$ & $\leftarrow$ \\
$T_{DIB}$ & Dayside integ. brightness temp. [K] & 1479 $\pm$ 13 K & $\leftarrow$ & $\leftarrow$ \\ 
$T_{NIB}$ & Nightside integ. brightness temp. [K] & 755 $\pm$ 46 K & $\leftarrow$ & $\leftarrow$ \\ 
\bottomrule
\end{tabular}
\tablecomments{This fully-shared fitting scenario was our preferred method, and thus these values represent the best-fit values of this work. Values given only in one column with arrows in adjacent columns represent parameters that were shared between the three sets. These values represent the median of the resulting posterior distribution determined from our MCMC sampling. The upper and lower errors given represent the 16th and 84th percentiles of the posterior distribution, respectively.}
\end{table*}


\subsection{Individual and Semi-Shared Fits}
\label{subsec:indivs}

We found that the phase curves in the individual fit case and the semi-shared fit case were very consistent with one another. The best-fit parameter values and relevant derived properties are given in Table~\ref{tab:indivfitvals} for the individual fit case, and in Table~\ref{tab:semisharedfitvals} for the semi-shared fit case. Between each individually fit visit, the primary phase curve parameters ($c_1$ and $c_2$) all agree within 2.18$\sigma$. The 2014 individually fit phase curve does exhibit a non-negligible second harmonic term, with an amplitude that is approximately 11\% that of the fundamental term. This is similar to what \cite{Stev16} found for the 2014 data, as their second harmonic amplitude is 14\% that of the fundamental term. The 2019a and 2019b phase curves, however, do not show significant second harmonic terms. The semi-shared phase curves are in similar consistency with one another. Between each individual phase curve and its corresponding semi-shared fit phase curve, the parameters are all consistent within 1$\sigma$. 

\begin{figure*}
    \centering
    \includegraphics[width=\textwidth]{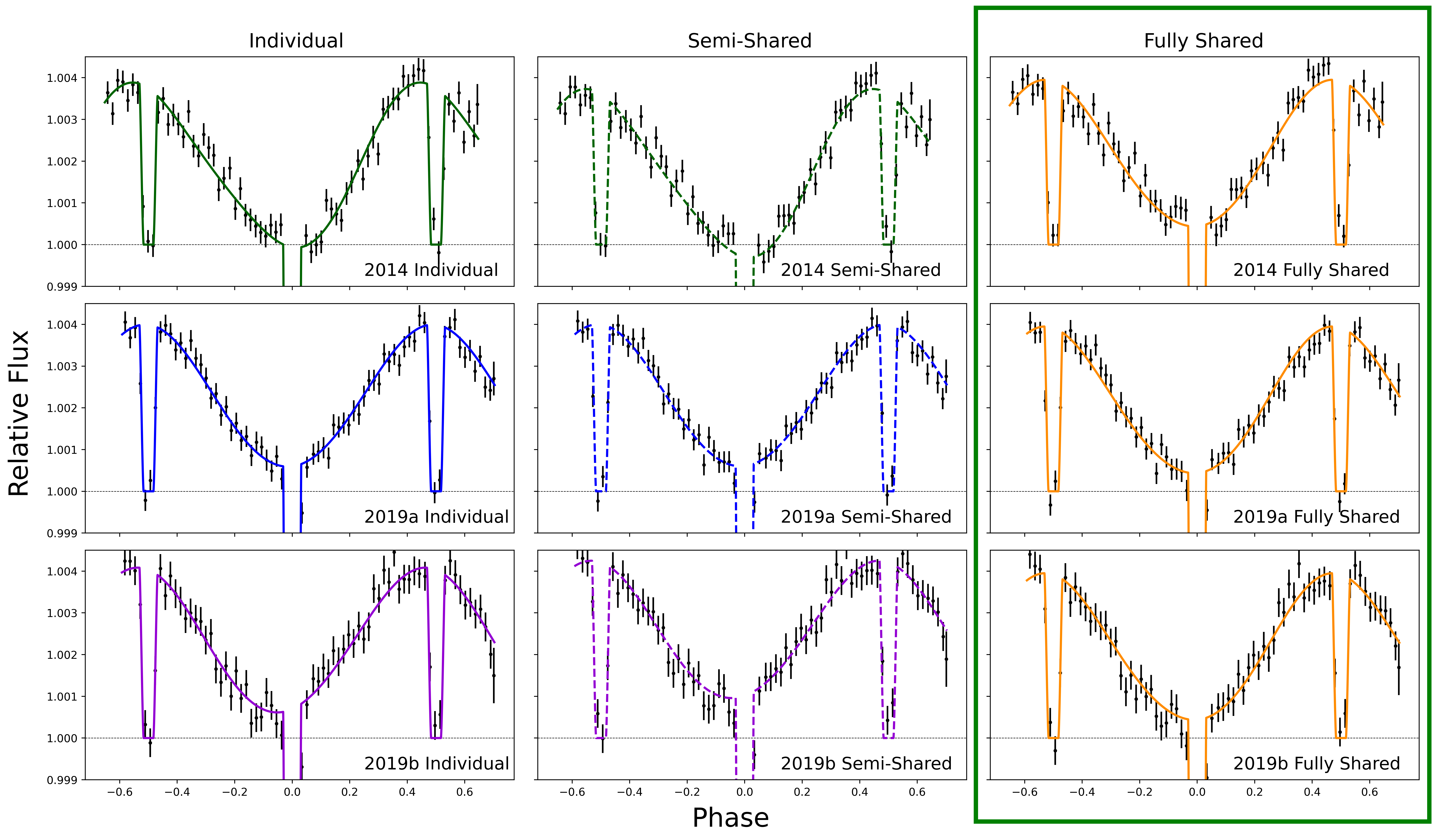}
    \caption{A plot of all light curves fit as part of this work. Columns distinguish different fitting methods -- leftmost are the individual fits, middle are the semi-shared fits, and rightmost are the fully shared fits. Rows distinguish different visits -- top being 2014, middle being 2019a, and bottom being 2019b. Our BIC analysis preferred the fully shared fit, meaning all three visits are best fit by a single, shared light curve.}
    \label{fig:sys_all}
\end{figure*}

\begin{figure*}
    \centering
    \includegraphics[width=\textwidth]{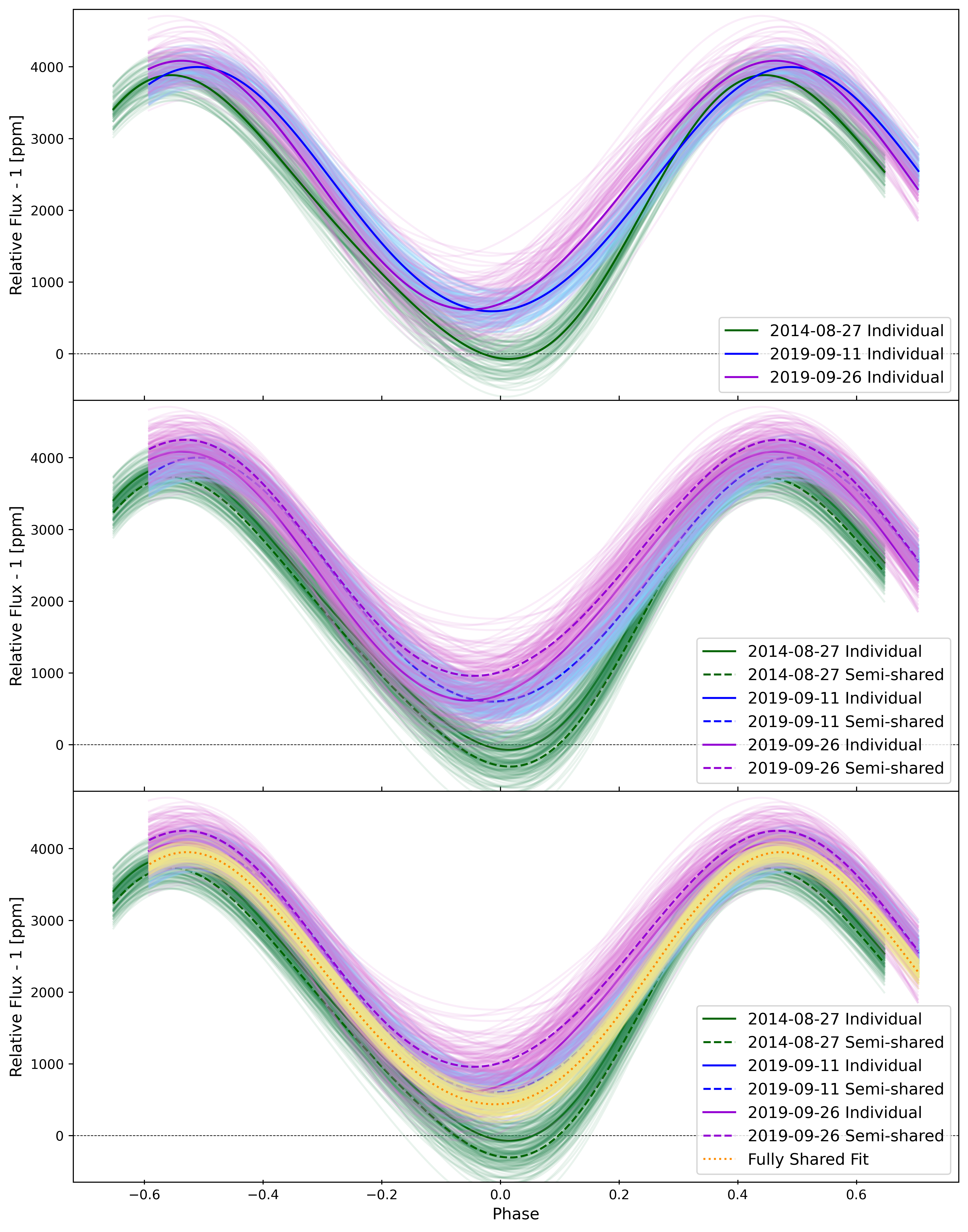}
    \caption{Overlays of the fitted phase curves from this work. Solid lines are the best fit, while the transparent lines represent the 1$\sigma$ uncertainty associated with each curve. 
    \textbf{Top:} the individually fit phase curves.
    \textbf{Middle:} the semi-shared fit phase curves overlayed with the individually fit phase curves.
    \textbf{Bottom:} all fit phase curves, including the fully shared fit phase curve which we deem representative of all three visits. 
    }
    \label{fig:pccom_combined}
\end{figure*}

We plot the full individually fit light curves in the left-hand column of Figure~\ref{fig:sys_all} and show just their phase curve components in the top panel of Figure~\ref{fig:pccom_combined}. Likewise, the semi-shared fit light curves are plotted in the middle column of Figure~\ref{fig:sys_all} and their phase curve components in the middle panel of Figure~\ref{fig:pccom_combined}. 

The best-fit R$_p$/R$_\star$ values were all consistent within 1.14$\sigma$, while the other planetary and orbital parameters were consistent within 0.3$\sigma$ between each individually fit visit. In addition, they are all within 1.09$\sigma$ of the semi-shared fit planetary and orbital parameters. As mentioned, all three visits used the same set of these parameters in the semi-shared fit case. The fact that the orbital parameters are all consistent between visits, even when fit entirely individually, indicates that our fits are self-consistent.                                         

When focusing on the phase curve, our fits of the 2014 visit do seem different from the 2019 visits at first glance. The individually fit 2014 visit's minimum phase curve value is -77 $\pm$ 200 ppm, which is lower than the 2019 visits' values of 568 $\pm$ 153 ppm and 583 $\pm$ 330, respectively. The 2014 visit also exhibits a much larger hotspot offset of -19.6$^\circ$ $\pm$ 2.5$^\circ$ compared to -4.2$^\circ$ $\pm$ 2.5$^\circ$ for the 2019a visit and -11.4$^\circ$ $\pm$ 6.8$^\circ$ for the 2019b visit, in the individually fit cases. The discrepancy between the minimum phase curve values is not statistically significant, as they only differ to 2.6$\sigma$. The hotspot offsets differ by 4.3$\sigma$, which is significant. However, we believe that these discrepancies are the result of strong intrapixel sensitivity effects in the 2014 data, rather than astrophysical variability. We checked for excess red noise left over in the detrended data by plotting the standard deviations of residuals as a function of temporal bin sizes for each visit's light curve, shown in Figure~\ref{fig:rednoisechecks} of the Appendix. We find that each visit's residuals roughly follow the 1/$\sqrt{N}$ relation expected from pure white noise, and that the 2014 visit does not appear to have significant excess noise compared to our 2019 visits. As we will discuss further in Section~\ref{subsec:discussion_comps}, other works have reanalyzed the 2014 data with different approaches to correcting for the subpixel sensitivity variations and found values that more closely match our results for the 2019 data \citep{Mendonca2018a, Morello2019, May2020, bell2021}.

\begin{table*}
\caption{Best-fit Values when Fitting the Three Data Sets Individually. }
\centering
\label{tab:indivfitvals}
\hspace*{-2.5cm}\begin{tabular}{lcccccl}\toprule 
Parameter & Units & 2014 & 2019a & 2019b \\\midrule  
Fitting Parameters & & & &  \\
$t_c$ - 2456612 & Previous transit time [BJD$_{\text{TDB}}$] & 0.41608 $\pm$ 4 $\times$ $10^{-5}$ & 0.41608 $\pm$ 4.3 $\times$ $10^{-5}$ & 0.41607 $\pm$ 4 $\times$ $10^{-5}$ \\
$\log_{10} \left(P\right)$ & log Orbital period [days] & -0.08965628 $\pm$ 2 $\times$ $10^{-8}$ & -0.08965629 $\pm$ 8 $\times$ $10^{-8}$ & -0.08965628 $\pm$ 2 $\times$ $10^{-8}$ \\
$\log_{10} \left(a / R_\star \right)$ & log scaled semi-major axis & 0.687 $\pm$ 0.001 & 0.687 $\pm$ 0.001 & 0.687 $\pm$ 0.001  \\
$\cos ~i$  & Cosine of inclination angle & 0.1368 $\pm$ 0.0009 & 0.1370 $\pm$ 0.0008 & 0.1373 $\pm$ 0.0009  \\
$R_p / R_\star$ & Radius ratio & 0.1587 $\pm$ 0.0006 & 0.1576 $\pm$ 0.0006 & 0.1574 $\pm$ 0.0009 \\
$c_1$ & 1st harmonic amplitude & 0.00193 $\pm$ 0.00008 & 0.00170 $\pm$ 0.00006 & 0.00173 $\pm$ 0.00014 \\
$c_2$ & 1st harmonic offset [deg] & 8.7 $\pm$ 1.9 & 4.6 $\pm$ 2.1  & 15.6 $\pm$ 5.6 \\
$c_3$ & 2nd harmonic amplitude & 2.3 $\times$ $10^{-4}$ $\pm$ 5 $\times$ $10^{-5}$ & -5 $\times 10^{-5}$ $\pm$ 6 $\times 10^{-5}$ & -6.2 $\times 10^{-5} \pm$ 1 $\times$ $10^{-4}$ \\
$c_4$ & 2nd harmonic offset [deg] & -88.0 $\pm$ 14.4  & -0.6 $\pm$ 53 & -11.8 $\pm$ 75.3 \\
$f_0$ & Phase curve vertical offset & 0.0018 $\pm$ 0.0001 & 0.0022 $\pm$ 0.0001 & 0.00230 $\pm$ 0.00022 \\
$s_t$ & Linear ramp slope & -0.00086 $\pm$ 0.00014 & -0.00043 $\pm$ 0.00013 & -0.0012 $\pm$ 0.0008 \\
$s_x$ & x-FWHM ramp slope & -0.023 $\pm$ 0.009 & -0.022 $\pm$ 0.004 & 0.002 $\pm$ 0.002 \\
$s_y$ & y-FWHM ramp slope & 0.039 $\pm$ 0.009 & 0.019 $\pm$ 0.005 & -0.007 $\pm$ 0.007 \\
Derived Parameters & & & & \\
$t_c$ & Current transit time [BJD$_{\text{TDB}}$] & 2456897.13201 $\pm$ 0.00004 & 2458738.83735 $\pm$ 0.00009 & 2458753.47983 $\pm$ 0.00010  \\
$t_s$ & Current eclipse time [BJD$_{\text{TDB}}$] & 2456897.53892 $\pm$ 0.00004 & 2458739.24426 $\pm$ 0.00009 & 2458753.88674 $\pm$ 0.00010 \\
P & Orbital Period [days] & 0.81347410 $\pm$ 5 $\times$ $10^{-8}$ & 0.81347406 $\pm$ 3 $\times$ $10^{-8}$ & 0.81347409 $\pm$ 3 $\times$ $10^{-8}$ \\
$a / R_\star$ & Scaled semi-major axis & 4.86 $\pm$ 0.02 & 4.86 $\pm$ 0.02 & 4.86 $\pm$ 0.02 \\
$i$ & Inclination angle [deg] & 82.126 $\pm$ 0.001 & 82.109 $\pm$ 0.001 & 82.131 $\pm$ 0.001 \\
$F_{\text{min}}$ & Minimum phase curve flux [ppm] & -77 $\pm$ 200 & 568 $\pm$ 153  & 583 $\pm$ 330 \\
$\delta_{\text{eclipse}}$ & Secondary eclipse depth [ppm] & 3742 $\pm$ 144 & 3984 $\pm$ 131 & 4012 $\pm$ 190  \\
$\Phi$ & Hotspot offset [degrees] & -19.6 $\pm$ 2.5 & -4.2 $\pm$ 2.5 & -11.4 $\pm$ 6.8 \\ 
$A_{P2P}$ & Peak-to-peak amplitude [ppm] & 3953 $\pm$ 179 & 3407 $\pm$ 112 & 3486 $\pm$ 284  \\\bottomrule 
\end{tabular}
\tablecomments{These values represent the median of the resulting posterior distribution determined from our MCMC sampling. The upper and lower errors given represent the 16th and 84th percentiles of the posterior distribution, respectively.}
\end{table*}

\begin{table*}
\caption{Best-fit Values when Fitting the Three Data Sets Simultaneously with Shared Planetary and Orbital Parameters but Individual Phase Curve and Systematic Parameters.} 
\centering
\label{tab:semisharedfitvals}
\hspace*{-2.7cm}\begin{tabular}{lcccccl}\toprule 
Parameter & Units & 2014 & 2019a & 2019b \\ \midrule 
Fitting Parameters & & & &  \\
$t_c$  & Previous transit time [BJD$_{\text{TDB}}$] & 2456612.41608 $\pm$ 4 $\times$ $10^{-5}$ & $\leftarrow$ & $\leftarrow$ \\
$\log_{10} \left(P\right)$ & log Orbital period [days] &  -0.08965628 $\pm$ 1 $\times$ $10^{-8}$ & $\leftarrow$ & $\leftarrow$ \\
$\log_{10} \left(a / R_\star \right)$ & log scaled semi-major axis & 0.687 $\pm$ 0.001 & $\leftarrow$ & $\leftarrow$  \\
$\cos ~i$  & Cosine of inclination angle & 0.1368 $\pm$ 0.0008 & $\leftarrow$ & $\leftarrow$  \\
$R_p / R_\star$ & Radius ratio & 0.1578 $\pm$ 0.0004 & $\leftarrow$ & $\leftarrow$ \\
$c_1$ & First harmonic amplitude & 0.00196 $\pm$ 0.00009 & 0.00170 $\pm$ 0.00006 & 0.00164 $\pm$ 0.00014 \\
$c_2$ & First harmonic offset [deg] & 7.7 $\pm$ 1.9 & 4.4 $\pm$ 2.1 & 13.6 $\pm$ 5.8 \\
$c_3$ & Second harmonic amplitude &  2.3 $\times$ $10^{-4}$ $\pm$ 5 $\times$ $10^{-5}$ & -6 $\times$ $10^{-5} \pm$ 6 $\times$ $10^{-5}$ & -7 $\times$ $10^{-5} \pm$ 1 $\times$ $10^{-4}$ \\
$c_4$ & Second harmonic offset [deg] & -86.6 $\pm$ 14.1 & 0.8 $\pm$ 48.2 & -0.2 $\pm$ 65.9 \\
$f_0$ & Phase curve vertical offset & 0.0017 $\pm$ 0.0001 & 0.0022 $\pm$ 0.0001 & 0.0025 $\pm$ 0.0002 \\
$s_t$ & Linear ramp slope & -0.0009 $\pm$ 0.0001 &  -0.0004 $\pm$ 0.0001 &  -0.0035 $\pm$ 0.0005 \\
$s_x$ & x-FWHM ramp slope & -0.019 $\pm$ 0.009 & -0.021 $\pm$ 0.006 & 0.001 $\pm$ 0.002 \\
$s_y$ & y-FWHM ramp slope & 0.019 $\pm$ 0.010 & 0.016 $\pm$ 0.010 & -0.002 $\pm$ 0.004 \\
Derived Parameters & & & & \\
$t_c$ & Current transit time [BJD$_{\text{TDB}}$] & 2456897.13201 $\pm$ 0.00004 & 2458738.83732 $\pm$ 0.00008 & 2458753.47985 $\pm$ 0.00008  \\
$t_s$ & Current eclipse time [BJD$_{\text{TDB}}$] & 2456897.53892 $\pm$ 0.00004 & 2458739.24423 $\pm$ 0.00008 & 2458753.88676 $\pm$ 0.00008 \\
P & Orbital Period [days] & 0.81347410 $\pm$ 3 $\times$ $10^{-8}$ & $\leftarrow$ & $\leftarrow$ \\
$a / R_\star$ & Scaled semi-major axis & 4.86 $\pm$ 0.01 & $\leftarrow$ & $\leftarrow$ \\
$i$ & Inclination angle [deg] & 82.132 $\pm$ 0.001 & $\leftarrow$ & $\leftarrow$ \\
$F_{\text{min}}$ & Minimum phase curve flux [ppm] & -298 $\pm$ 214 & 553 $\pm$ 152 & 891 $\pm$ 333 \\
$\delta_{\text{eclipse}}$ & Secondary eclipse depth [ppm] & 3601 $\pm$ 140 & 3988 $\pm$ 135 & 4198 $\pm$ 190 \\
$\Phi$ & Hotspot offset [degrees] & -18.9 $\pm$ 2.5 & -3.6 $\pm$ 2.5 & -12.2 $\pm$ 6.9 \\ 
$A_{P2P}$ & Peak-to-peak amplitude [ppm] & 4041 $\pm$ 189 & 3430 $\pm$ 115 & 3302 $\pm$ 279 \\\bottomrule 
\end{tabular}
\tablecomments{Values given only in one column with arrows in adjacent columns represent parameters that were shared between the three sets. These values represent the median of the resulting posterior distribution determined from our MCMC sampling. The upper and lower errors given represent the 16th and 84th percentiles of the posterior distribution, respectively.}
\end{table*}

\subsection{Comparison of Fits}
\label{subsec:fitcomps}

The three fitting scenarios used span the range of potential variation between visits. If there is significant variation, then one of the two scenarios where the three visits had separate phase curve parameters ought to be preferred. If there is no variation, then the fully shared scenario ought to be preferred. 

We compare the goodness-of-fit of each scenario using the BIC, and find that the BIC decisively prefers the fully shared fit. The semi-shared fit came second, with $\Delta |\text{BIC}| = -80$ relative to the absolute BIC of the fully shared fit, and the individual fits were least preferred with $\Delta |\text{BIC}| = -112$.
The fully shared fit phase curve, therefore, represents a single phase curve of the planet, whose emission does not strongly change with time. We plot the corresponding full astrophysical model along with the combined data from all three visits in Figure \ref{fig:sys_sharedfitalldata}. 

\section{Weather Models}
\label{sec:atmomodel}

\subsection{GCM Setup}
\label{subsec:modelsetup}

To interpret the observed phase curves, we simulated atmospheric conditions on WASP-43b using a 3D general circulation model (GCM). Our GCM used a double-gray radiative transfer scheme. Specifically, we used the model of \cite{rauscher+2010,rauscher+2012}, adapted from \cite{Hoskins1975}, with recent updates to account for radiative feedback and scattering from clouds \citep{Roman&Rauscher2017,Roman&Rauscher2019, roman2021clouds}. In particular, we used the cloud model of \cite{roman2021clouds} that includes up to 13 different cloud species for various vertical distributions. 

In the \cite{roman2021clouds} model, clouds are treated as purely temperature-dependent sources of opacity, with constant mixing ratios set by the assumed solar elemental abundances. The optical thicknesses of the clouds are determined by converting the relative molecular abundances (or partial pressures) of each species into particles with prescribed densities and radii, following \cite{roman2021clouds}.

We ran simulations for both clear atmospheres and two different sets of cloud compositions. Following \cite{roman2021clouds}, one cloud composition case includes 13 different species: KCl, ZnS, Na$_2$S, MnS, Cr$_2$O$_3$, SiO$_2$, Mg$_2$SiO$_4$, VO, Ni, Fe, Ca$_2$SiO$_4$, CaTiO$_2$, and Al$_2$O$_3$. We refer to this case using all 13 species as the ``all species" case. We also modeled a case in which only eight of these species are included, based on considerations of nucleation efficiency \citep{gao2020aerosol}. This compositional case lacks iron, nickel, and three sulfides, so the clouds are limited to KCl, Cr$_2$O$_3$, SiO$_2$, Mg$_2$SiO$_4$, VO, Ca$_2$SiO$_4$, CaTiO$_2$, and Al$_2$O$_3$. We refer to this case with only eight species as the ``nucleation limited" case. We refer the reader to \citep{roman2021clouds} for more details on the cloud properties and parameters for each species. In general, the silicates form thick, conservatively scattering clouds, while the Fe, Al$_2$O$_3$, and Cr$_2$O$_3$ form clouds of lower albedos that absorb starlight. Places where clouds overlap have mixed properties, weighted by the optical thickness of each species. 

For both cloud composition scenarios, we explored the observational consequences of variations in the cloud deck's vertical thickness.  While we expect the vertical extent of the clouds to be determined by a balance between vertical mixing and settling, along with other limiting processes, these processes are computationally expensive and poorly constrained both observationally and theoretically. Rather than attempt to model these cloud physics explicitly, we therefore defined the maximum vertical thickness, relative to the cloud base pressure, after which the cloud is truncated, and run simulations for a range of values. We choose values at five-layer intervals, ranging from 5 to 45 layers of our 50-layer model. Each cloud species is allowed to have its own individual vertical extent, though all species are subject to the same maximum limit. This effectively mimics a range of vertical mixing strengths and amounts to nine simulations for each of our two cloud composition scenarios. Each 5-layer interval roughly corresponds to one scale height, and we describe each model's maximum vertical thickness in terms of the number of scale heights hereafter. 

Our model assumes a constant optical depth per bar of pressure, which results in a constant aerosol mixing ratio for each species. As a result, vertically thicker clouds are correspondingly optically thicker in our model. This directly ties the emission characteristics in our cloudy atmosphere to the vertical extent of the clouds, so that the vertical thickness directly traces the optical thickness of the cloud above the thermal photosphere. In reality, however, optical depth can also be set by the concentration of cloud particles.

Using the GCM and planetary parameters listed in Table \ref{tab:atmomodparams}, we began our simulations with clear skies, no winds, and no horizontal temperature gradients. We ran the simulations for 3500 planetary orbits, assuming tidal synchronization. We allowed the winds, temperatures, and clouds to evolve over the course of the simulation. The results from the GCMs were recorded at 100 day intervals to track possible temporal variability. Additional details of the GCM dynamics can be found in \cite{rauscher+2012}, \cite{Roman&Rauscher2019}, and \cite{roman2021clouds}. Once the GCM runs completed, we reduced them to the corresponding emission phase curves at 4.5~$\mu$m. 

\subsection{Model Phase Curve Extraction}
\label{subsec:phasecurvemodeling}
In order to generate phase curves from the GCM, we followed the routine outlined by \cite{Malsky2021}. First, we remapped the GCM from constant pressure levels to 250 constant altitude layers by using vertical hydrostatic equilibrium, as assumed in the GCM. Next, we recovered the locations and properties of each cloud species, for each layer, by using the same criteria as within the GCM. We calculated the particle size, wavelength-dependent extinction efficiency, asymmetry parameter, and single scattering albedo for each cloud species on a 50x50 wavelength and pressure grid following the Mie scattering code from \cite{Wolf2004} and the \texttt{Rosseland Clouds} software package\footnote{\url{https://github.com/ELeeAstro/Rosseland_Clouds}}. Lastly, we used the vertical gradient from \cite{roman2021clouds} to relate the atmospheric pressure to the predicted particle sizes for the radiatively active clouds. At pressures of less than 10 mbar we assumed particles with radii of 0.1 $\mu$m, with an exponential increase to particle sizes of approximately 80 $\mu$m at 100 bar. 

We calculated simulated phase curves from the GCM models using the output 3D atmospheric structure. Using a line-by-line ray-tracing radiative transfer routine that is set up to calculate the intensity toward the observer from each line-of-sight path through the atmosphere \citep[e.g][]{zhang2017}, we calculated the emission spectrum from the planet, for viewing geometries throughout its orbit. We assumed solar abundance (consistent with the GCMs) and recovered the cloud properties as described above. We then integrated the spectrum at each phase over the Spitzer 4.5 micron filter response function to calculate the photometric flux from the planet at each phase. To convert these integrated emission phase curves into relative flux units, we divided by the band-integrated stellar flux computed from a BT-Settl model stellar spectrum of the star WASP-43.

\begin{table*}[th!]
\centering
\caption{GCM Parameters}
\label{tab:atmomodparams}
\begin{tabular}{lccccl} \toprule
 Parameter                  & Value              & Units                & Comment \\\midrule 
\textit{Orbital/Dynamical}  & & & \\
Radius of the planet, $R_p$ & 7.407 $\times$ $10^7$ & m & 1.036 $R_{\text{Jupiter}}$ \\
Gravitational acceleration, $g$ & 47.02  &m s$^{-2}$ &  \\
Rotation rate, $\Omega$ & 8.94 $\times$ $10^{-5}$ & s$^{-1}$ & tidally synchronized, 0.813 day orbit\\\\
\it{    Radiative Transfer}\\
Specific gas constant, $\mathcal{R}$ & 3523 & J kg$^{-1}$ K$^{-1}$ & assumed $H_2$-rich\\
Ratio of gas constant to heat capacity, $\mathcal{R}/c_P$ & 0.286 & -- & assumed diatomic \\
Incident stellar flux, $F_{\downarrow \mathrm{vis}}$ & 9.78 $\times 10^7$ & W m$^{-2}$ \\
Internal heat flux, $F_{\uparrow \mathrm{IR}, \mathrm{int}}$ & 3544 & W m$^{-2}$  & from modeled 1-D T-profile\\
Gaseous visible absorption coefficient, $\kappa_{\mathrm{vis}}$  & 4.00 $\times$ $10^{-3}$  & cm$^2$ g$^{-1}$ & constant, from modeled 1-D T-profile \\
Gaseous visible scattering coefficient, $\kappa_{\mathrm{Ray}}$  & 1.72 $\times$ $10^{-4}$  & cm$^2$ g$^{-1}$ & Rayleigh scattering for a spherical albedo of $\sim$0.15 \\
Gaseous infrared absorption coefficient, $\kappa_{\mathrm{IR}}$  & 1.00 $\times$ $10^{-2}$  & cm$^2$ g$^{-1}$  & constant, from modeled 1-D T-profile\\\\
\it{    Model Resolution}\\
Vertical layers & 50 & --\\
Bottom of modeling domain pressure &  $\sim$100 & bar \\
Top of modeling domain pressure & 5.7 $\times$ $10^{-5}$ & bar\\
Horizontal resolution & T31 & -- & corresponds to  $\sim$48 lat $\times$  $\sim$96 lon\\
Dynamical temporal resolution & 4800 & time steps/day & \\
Radiative transfer temporal resolution  & 1200 & time steps/day & heating rates updated every four time steps\\
Simulated time & 3500 & planet days & 3500 revolutions\\\bottomrule
\end{tabular}
\end{table*}

\subsection{Variability Analysis}
\label{subsec:modelresults}

Variability due to cloud-driven weather may manifest in multiple ways. For example, the cloud thickness may remain constant but the spatial (i.e. latitudinal and longitudinal) distribution of the clouds at any given phase may change \citep[e.g.][]{lines2018}. Alternatively, the vertical cloud thickness may change, but the spatial distribution remains roughly constant. Since vertical cloud thickness is a freely prescribed parameter in our models, we are able to investigate this second scenario directly. We remind the reader that by the ``thickness" of each model, we are referring to the maximum vertical extent beyond which the clouds were truncated. By comparing our observations to the extracted phase curves from models of different thicknesses, we seek to constrain the extent to which the thickness of WASP-43b's clouds may have changed between observations while still remaining consistent with the data.

We first checked for temporal variability within each GCM model, by comparing the extracted phase curve at different output times after the initial settling period. Such variability could be caused by changes in the spatial distribution of clouds. We do not see any significant temporal variations in these phase curves, implying that there is likely no significant spatial variation in the cloud coverage at scales larger than the model's spatial resolution of T31 (Table~\ref{tab:atmomodparams}). 

\begin{figure*}
    \centering
    \includegraphics[width=\textwidth]{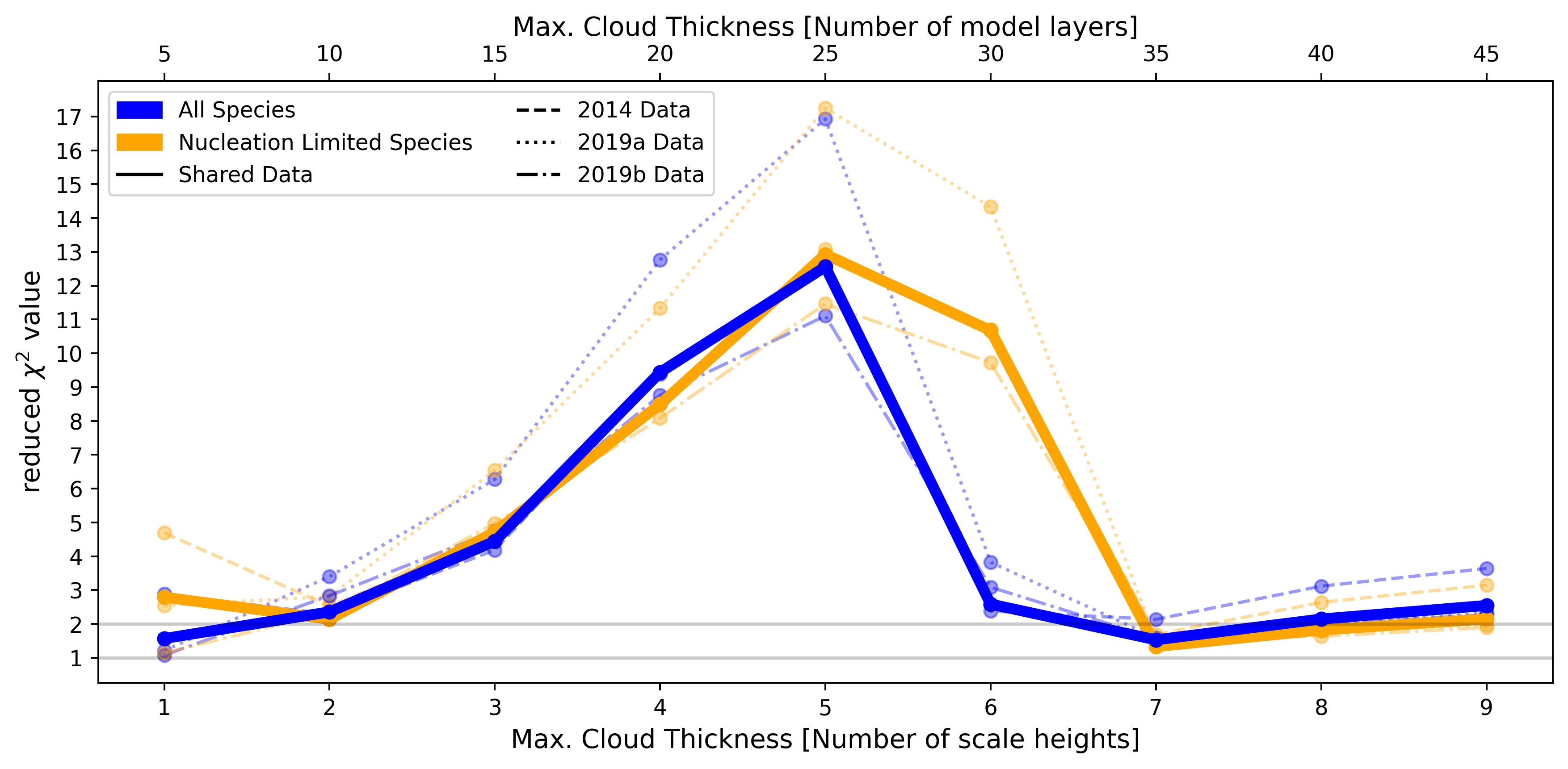}
    \caption{Reduced-$\chi^2$ values between our observed phase curve data and each cloudy GCM phase curve. The dashed, dotted, and dot-dashed lines show this comparison for the 2014, 2019a, and 2019b visits individually, respectively. The thick, solid lines show this comparison for all three observing visits combined. Horizontal lines are drawn at reduced-$\chi^2$ = 1 and 2 for visual reference. Our observations strongly reject the intermediate thickness models. We find that no model fits the data extremely well, as the minimum reduced-$\chi^2$ = 1.3 for the seven scale height nucleation limited model. However, we find that our observations are reasonably consistent with both the thinnest models (of 1 - 2 scale height maximum thicknesses) and the thickest models (7 - 9 scale height layer maximum thicknesses).
    }
    \label{fig:sigmatestresults}
\end{figure*}

We also compared our observations to various phase curves extracted from GCMs with different cloud thickness values. We computed the reduced-$\chi^2$ statistic between each model phase curve and the detrended data for each individual visit, as well as the three visits combined -- excluding data points within transit and eclipse. Figure~\ref{fig:sigmatestresults} shows the reduced-$\chi^2$ values for each comparison as a function of the model's maximum cloud thickness. 

\begin{figure*}
    \centering
    \includegraphics[width=\textwidth]{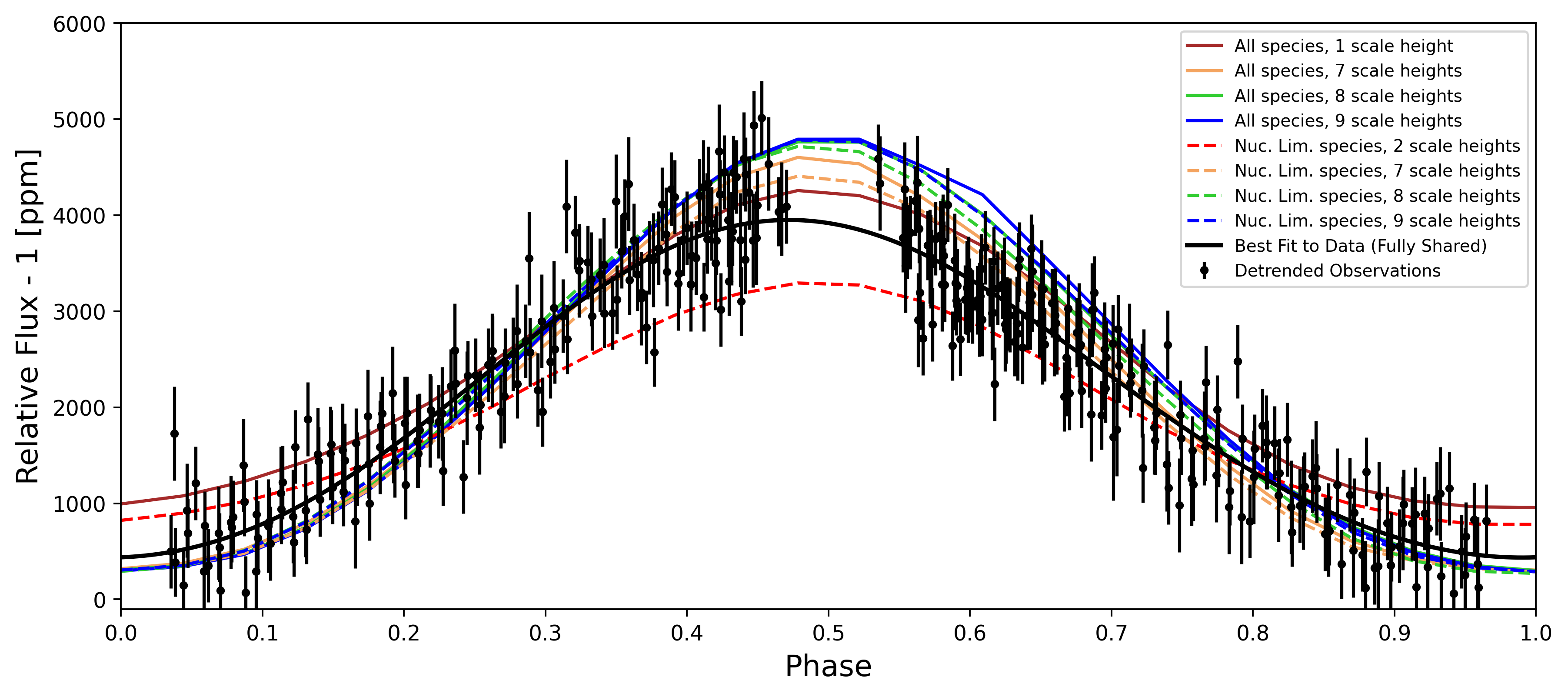}
    \caption{ Plot of the cloudy GCM phase curves that are consistent with our combined observations by $\chi^2_{reduced} < $ 2.2. Models with all cloud species are shown as colored solid lines, while those with nucleation limited species are the colored dotted lines. Also shown are the detrended observed data, combined from the 2014, 2019a, and 2019b visits, as fit by our fully shared model. Points within transit and eclipse are removed. The solid black line is the fully shared fit phase curve, which was best fit to the data as described in Section~\ref{sec:results}. }
    \label{fig:obs_closemodel_curves}
\end{figure*}

As shown in Figure~\ref{fig:sigmatestresults}, no single GCM output perfectly fits the data. We can, however, rule out a wide range of possible cloud thickness models, as all models with thicknesses of 3 to 6 scale heights are strongly rejected with reduced-$\chi^2$ values ranging from 2.6 - 12. The remaining models with very thin clouds (1 to 2 scale heights) and very thick clouds (7 to 9 scale heights) are marginally consistent with the detrended observations. 
Figure~\ref{fig:obs_closemodel_curves} shows these remaining model phase curves  overplotted with our observations, with all three visits combined, and the fully shared phase curve from section~\ref{subsec:fullyshared} that was best fit to these data.

Our observations strongly reject intermediate thicknesses. But, our observations cannot distinguish between the very thin models and the very thick models, as the data are statistically fit equally well by both. This statistical degeneracy is, in part, due to the inability to collect phase curve measurements at transit or eclipse, when we would see only the nightside or dayside, respectively. Rather, the majority of our points of comparison span phases where we see the terminator regions, where the model phase curves are much more similar than they are at day or nightside phases. It does, however, seem like our thin and thick models make fairly similar predictions for the dayside emission. For example, the peak emissions of the one scale height phase curve and the seven scale height phase curve only differ by $\sim$ 1 data error bar. Cloud feedback provides an explanation for why such different thicknesses could lead to similar phase curves. The thin cloud model has clouds condensing at or below the photosphere leading to a clear, hot atmosphere with strong emission. At intermediate thicknesses, reflective silicate clouds begin condensing above the photosphere which reduces absorption, cools the atmosphere, and lowers emission. The thickest models, however, have built up sufficient vertical extent that highly absorbing clouds with high condensation temperatures finally peak above the photosphere, which re-increases absorption on the dayside, warms the local atmosphere, and re-increases dayside emission. This effect of increased emission due to cloud feedback has also been seen by \cite{Roman&Rauscher2019}. It is likely also true that, since our models' clouds are allowed to thin below their maximum prescribed thickness due to evaporation, some of the dayside clouds in our thickest models have evaporated, which will also affect the dayside emission.


The lower dayside emission predicted by the thin one scale height model of 4256 ppm is closest to our observed maximum relative flux of 3936 $\pm$ 84 ppm. Conversely, the lower nightside emission predicted by the thick models of $\sim$300 ppm best match our observed minimum relative flux of 425 $\pm$ 105 ppm. We believe that, in reality, WASP-43b has a complicated combination of thick nightside clouds and relatively thinner dayside clouds which our models cannot accurately disentangle. 

Since neither of the seven, eight, and nine scale height models can be strongly rejected from one another, this thicknesses range of three scale heights places an upper limit on how much the maximum cloud thickness on WASP-43b may have varied between visits. Combining this with the lack of spatial variability seen, our observations suggest that WASP-43b's clouds are stable in terms of their vertical and spatial distributions over timescales as long as several years.


\section{Predictions for JWST Observations}
\label{sec:jwst}

\textit{JWST} has successfully launched, and has shown great promise to deliver years of exciting infrared science as a successor to the \textit{Spitzer Space Telescope}. In particular, \textit{JWST} is well equipped to study atmospheric variability on exoplanets. \textit{JWST}'s unprecedented precision should be able to probe smaller phase curve amplitude variations, and its wider wavelength range may be able to probe silicate cloud spectral features in the 5 - 10~$\mu$m range. During the review process of this work, \textit{JWST} observed a single spectroscopic phase curve of WASP-43b using \textit{JWST}/MIRI LRS for the \textit{JWST} Transiting Exoplanet Community Early Release Science Program (hereafter ERS) \citep{bean2018ers}. Here, we predict how well such an observation could further constrain the nightside cloud thickness, and the level of thickness variations in WASP-43b's atmosphere. 

MIRI LRS covers a wide wavelength range of 5 - 12~$\mu$m, but achieves its highest precision around 5 - 6~$\mu$m \citep[][and our PandExo calculations]{glasse2015miri, Kendrew2015mirilrs}. For simplicity, we only consider this 5 - 6~$\mu$m bandpass. Following the same method as in Section~\ref{subsec:phasecurvemodeling}, we extracted cloudy phase curves in this bandpass from our GCMs. 

We then simulated MIRI LRS observations of each model phase curve. We used 7172 integrations of 9.86 seconds each, as computed by PandExo \citep{Batalha2017} for a full-orbit observation of WASP-43b. This PandExo estimate is slightly shorter than the ERS plan to use 8595 integrations of 10.3 seconds each, since they include a longer baseline to test for systematics. We scattered data points around each model phase curve at a 9.86 second cadence, assuming Gaussian noise. We set the 1$\sigma$ noise level equal to the per-integration flux uncertainty of 603 ppm, which we calculated from our PandExo simulation. 

\begin{figure*}
    \centering
    \includegraphics[width=\textwidth]{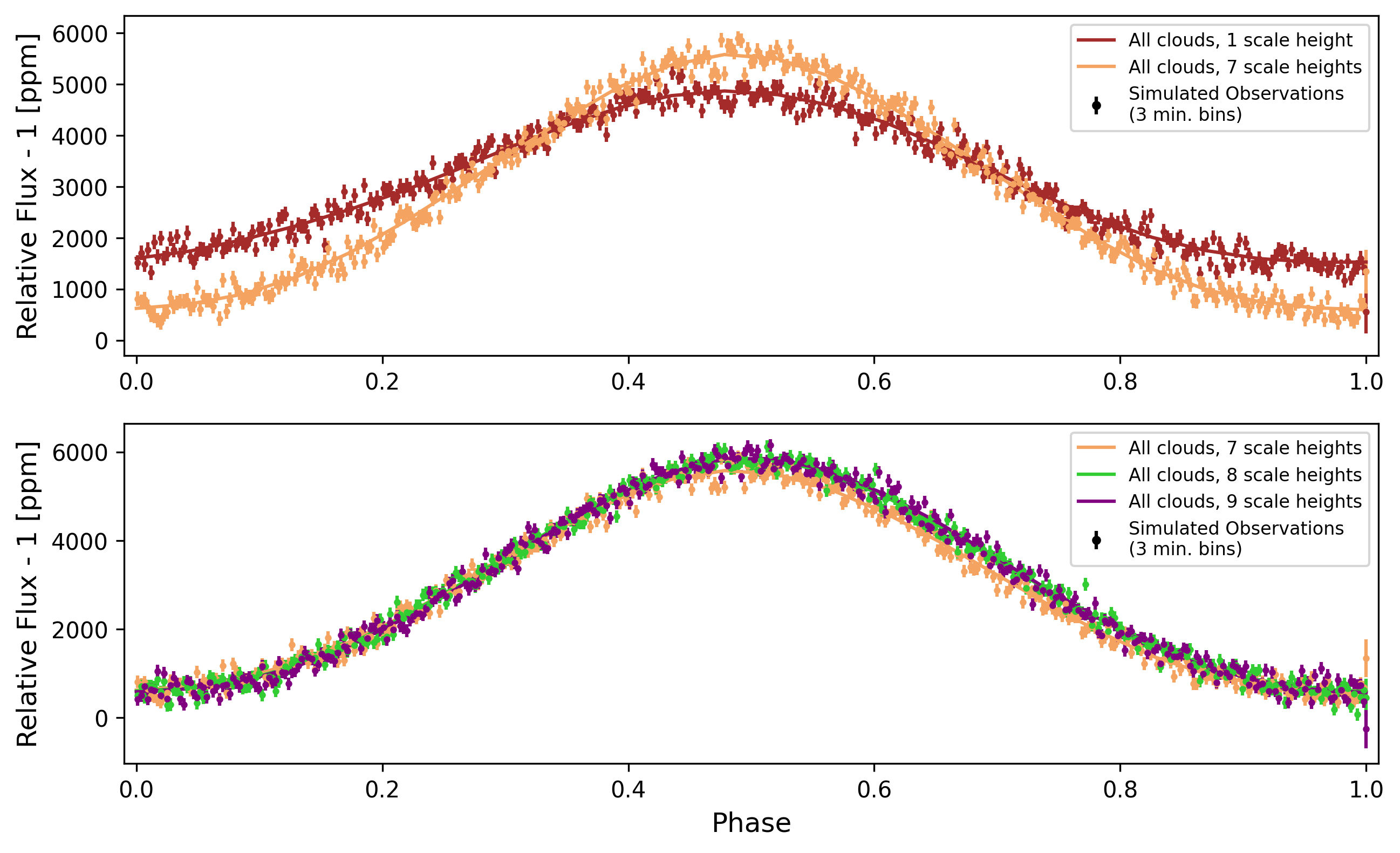}
    \caption{
    Simulated cloudy phase curves from our GCM integrated over a 5 - 6~$\mu$m bandpass. The points are simulated observations in this bandpass by \textit{JWST}/MIRI LRS, as a prediction for what the upcoming \textit{JWST} Transiting Exoplanet Community Early Release Science Program observations of WASP-43b \citep{bean2018ers} may achieve.
    \textbf{Top:}
     here, we focus only on the one and seven scale height models as these represent the two thickness regimes that our Spitzer observations could not distinguish. Assuming one of these models as truth, these MIRI LRS observations would be able to rule out the other model at very high significance. This suggests that MIRI LRS observations will be able to measure accurately WASP-43b's nightside emission and approximate cloud thickness. 
    \textbf{Bottom:} 
    here, we show the 7 - 9 scale height models, since our Spitzer observations could not significantly rule out any of these models, thus placing an upper limit on the possible thickness variation. We find that, in this 5 - 6~$\mu$m bandpass, there is no significant difference between the three models at any phase. Therefore, repeated MIRI LRS observations would likely not be able to place an improved upper limit on the thickness variations.
    }
    \label{fig:miricurves}
\end{figure*}

First, we tested whether these faux MIRI LRS observations could significantly distinguish between the thin and thick models, which our \textit{Spitzer} data could not. For this test, we focused just on the one and seven scale height models with all cloud species. The top panel of Figure~\ref{fig:miricurves} shows these one and seven scale height model phase curves and their simulated observations binned into three minute intervals. To test how well we could rule one model out given the other as a truth, we computed the reduced-$\chi^2$ statistic between the simulated observed data of one model and the other model's phase curve. We find that we could distinguish each model from the other at high significance with $\chi^2_{reduced} >$  2.2. 

As mentioned in Section~\ref{subsec:modelresults}, a primary difference between these thin and thick models is in their prediction of the nightside emission. We repeated the above reduced-$\chi^2$ comparison using only points at phases within $\pm$ 0.1 of phase 0 or phase 1 (phase 0 and 1 are equivalent). With this subset of points, we find that we could distinguish the nightside predictions at much stronger significance compared to the full-orbit data, of $\chi^2_{reduced} >$ 3.5. This suggests that the \textit{JWST}/MIRI LRS phase curves of WASP-43b being taken during the upcoming ERS observations will be able to place a strong constraint on WASP-43b's nightside cloud thickness. We note the caveat, however, that in our models there are other factors in addition to just the prescribed cloud thickness that contribute to the exact amount of emission at any given phase. 

Then, we tested whether these MIRI LRS observations, if repeated, could place a better limit on thickness variations than the three scale height upper limit made from our \textit{Spitzer} observations. The bottom panel of Figure~\ref{fig:miricurves} shows the 7 - 9 scale height model phase curves and their simulated observations, again binned to three minute intervals. We find that, in this 5-6~$\mu$m integrated bandpass, there is little difference between the models at any phase and the simulated MIRI LRS observations also cannot distinguish between them. Therefore, repeated MIRI LRS observations would likely not be able to place a better limit on the potential cloud thickness variation in WASP-43b's atmosphere.

\section{Discussion}
\label{sec:discussion}

\subsection{Reanalyses of WASP-43b's 4.5~micron phase curve}
\label{subsec:discussion_comps}

The results of \textit{Spitzer}/IRAC phase curve observations are generally sensitive to where the star landed on the detector, and the data reduction method used, due primarily to complicated intrapixel sensitivity variations \citep[e.g. see][]{May2020, May2022}. This systematic error is likely the source of the discrepancy between our individual and semi-shared fits of the 2014 data, and those of the 2019 data. 

The first major discrepancy between our 2014 and 2019 phase curves is the minimum phase curve value. In the individually fit case, we found a zero-consistent value of -77 $\pm$ 200 for the 2014 visit. The original analysis of the 2014 data also found a near-zero value of 19 ppm \citep{Stev16, May2020}, and the reanalysis of \cite{bell2021} also found a low value of $\approx$196 ppm. Other reanalyses of the 2014 data by \cite{Mendonca2018a}, \cite{Morello2019}, and \cite{May2020} find larger minimum phase curve values of 841 ppm, 300 $\pm$ 150 ppm, and 542 $\pm$ 124 ppm, respectively. These values more closely match what we found for our 2019 visits, of 568 $\pm$ 153 ppm and 583 $\pm$ 330 ppm, and suggest that WASP-43b's minimum phase curve value is indeed likely significantly nonzero as should be true for a nightside with a nonzero temperature. 

The other major discrepancy is in the hotspot offset. We find a hotspot offset of -19.6 $\pm$ 2.5 for the 2014 visit, similar to the -21.1$^\circ$ $\pm$ 1.8$^\circ$ value found by \cite{Stev16}. However, \cite{Mendonca2018a}, \cite{Morello2019}, \cite{May2020}, and \cite{bell2021} find values of -12$^\circ$ $\pm$ 3$^\circ$, -11.3$^\circ$ $\pm$ 2.1$^\circ$, -20.6$^\circ$ $\pm$ 2.0$^\circ$, and -20.4$^\circ$ $\pm$ 3.6$^\circ$ respectively. These values display significant scatter, and half are consistent with the hotspot offset value of -11.4$^\circ$ $\pm$ 6.8$^\circ$ that we found for our 2019b visit. These discrepancies in the hotspot offset are likely not astrophysical. As \cite{bell2021} discuss, the form of phase curve model chosen seems to drive scatter in the exact value of the hotspot offset. Analyses that use, and find contribution from, the second harmonic term tend to find a larger offset than those which use only a single, fundamental term. 

\subsection{Global cloud properties in hot Jupiter atmospheres}
\label{subsec:cloudiness_discussion}

When compared to our observations, neither our thinnest or thickest cloud models could be statistically rejected. As discussed in Section~\ref{subsec:modelresults}, though, the thinner model provided a closer prediction for the dayside emission to that of our fully shared phase curve, and same of the thick model for the nightside emission. The inability to reject either model is, in part, due to the limited resolution of our model and degeneracy in the dayside emission due to cloud feedback and evaporation. We believe that the most likely scenario is that WASP-43b has a combination of a cloudy nightside and a relatively cloud-free dayside. 

Our findings agree with the general picture of WASP-43b's atmosphere painted by previous observations. All observed phase curves of WASP-43b, with \textit{HST}/WFC3 \citep{stevenson2014} and \textit{Spitzer}/IRAC \citep{Stev16}\footnote{One of whose data sets we have reanalyzed in our work here}, have seen low relative flux on the nightside that cannot be recreated by cloud-free models \citep[e.g.][]{kataria2015}. Meanwhile, eclipse observations using \textit{HST}/UVIS \citep{fraine2021}, \textit{CHEOPS}, and \textit{TESS} \citep{scandariato2022} have found very low dayside albedos, suggesting a lack of clouds and hazes on the dayside. \cite{Helling20}'s models predicted significant formation of mineral clouds and photochemical hazes on WASP-43b's dayside, albeit much thinner than on the nightside. It may be the case that these dayside aerosols are sufficiently deep that they have hid from previous observations (see \cite{fraine2021} for a good discussion of this idea).

Nightside cloud formation is a common and generally undisputed model prediction. The formation and feedback effects of dayside clouds and hazes, however, are less understood. In fact, the similarity we found between the dayside emission of our thinnest and thickest models highlights the immense challenge of accurately modeling cloud formation in exoplanet atmospheres, and interpreting observations thereof. The microphysics involved in cloud formation are difficult to simulate on their own, and the problem becomes extremely complex if trying to include active formation along with effects like radiative feedback, scattering, and absorption in 3D geometries.  It is crucial to keep these complexities in mind when attempting to connect model results to spatially unresolved observations. 

Our observations and models, both in this work and in general, are resolution limited. Real atmospheres may, and likely do, vary at small spatial scales. Unless models use a very fine spatial grid, which would be difficult in terms of practicality and expense, it will be difficult to study these small scale variations. Additionally, with current observational capabilities, our data are typically limited to disk-integrated quantities so we are inherently insensitive to any atmospheric variations that happen to integrate out to the same observable value. This will likely remain an issue until more spatially resolved observations, such as the technique of eclipse mapping \citep[see, e.g.,][]{rauscher2007, dewit2012, majeau2012, rauscher2018, mansfield2020, challener2022}, become more common.

\subsection{Implications of No Variability}
\label{subsec:discussion_novar}

Between our 2019 observations, we place upper limits of 210 ppm or 5.0\% variation in the eclipse depth, and 128 ppm or 3.7\% variation in the peak-to-peak amplitude. These constraints are in line with the smaller scale atmospheric variability predicted by \cite{showman2009}, \cite{dobbsdixon2013}, \cite{komacek2020}, and \cite{menou2020}. We do not see the strong cloud-induced variability predicted in the models by \cite{lines2018}.

Our empirical limits on the variability in WASP-43b's phase curve are similar to those for HD 189733b \citep{agol2010, kilpatrick2020} and HD 209458b \citep{crossfield2012, kilpatrick2020}, which were also made using \textit{Spitzer}. Considered alone, these infrared observations would indicate that hot Jupiter atmospheres are quite stable in time. The significant variability observed at optical wavelengths by \cite{armstrong2016} and \cite{jackson2019} are in tension with this picture, assuming they are indeed the result of weather and not stellar contamination. Theory indeed predicts that the observable effect of weather is stronger at shorter wavelengths \citep{rauscher2008}. There may indeed be a transition from significant to insignificant variability as wavelength increases. In this case, our observations suggest that this transition lies below 4.5~\mum. Future higher-precision observations will enable finer constraints on the atmospheric variability.

\section{Summary \& Conclusions}
In this work, we analysed two new Spitzer/IRAC 4.5$\mu$m phase curves of the hot Jupiter WASP-43b, and reanalyzed the previous 4.5$\mu$m phase curve of this planet observed by \cite{Stev16}. Altogether, these three full-orbit observations offered three samples to test for variability in WASP-43b's atmosphere. 

We found that WASP-43b's three phase curves are best fit by a single, fully-shared model rather than three individual models. This shows that WASP-43b's phase curve does not change in time. The lack of observed variability in these Spitzer observations confirms theoretical predictions that weather-induced variability may be weak \citep{showman2009, dobbsdixon2013, komacek2020, menou2020}, and matches the lack of variability seen in longer wavelength Spitzer observations of other hot Jupiters \citep{agol2010, crossfield2012, kilpatrick2020}. 

To model our observations, we ran a suite of cloudy GCMs. Our models included active cloud formation and were run forward in time to capture potential variability. Each model had a different maximum allowed cloud thickness, beyond which clouds were truncated, to explore the effect of vertical thickness. Our models showed no time variability in their derived phase curves. We did some preliminary modeling to test how other properties vary within our models, such as the total integrated optical depth. This initial work showed a few percent variability in the total optical depth, but more detailed analysis is necessary before drawing any firm conclusions.
Furthermore, by comparison to our \textit{Spitzer}/IRAC observations, we found that the maximum cloud thickness in WASP-43b's atmosphere cannot be varying by more than three pressure scale heights over time. Together, these results suggest that WASP-43b's clouds are stable in their vertical and spatial extents for periods as long as several years.

Our observations were not able to rule out the thinnest cloud models, or the thickest cloud models. We found that the thick models best resembled WASP-43b's nightside, whereas the thin models best resembled the dayside. This suggests that WASP-43b likely has a cloudy nightside and a dayside which is relatively cloud-free at the pressures probed by the observations.

Lastly, we simulated phase curve observations of WASP-43b using JWST/MIRI LRS integrated over a 5-6~$\mu$m band to predict how the upcoming ERS observations of WASP-43b \citep{bean2018ers} may improve on our constraints made here with Spitzer. Our simulated MIRI LRS observations were able to distinguish the thin and thick cloud models significantly. This implies that the upcoming ERS observations with MIRI LRS will be able to measure precisely WASP-43b's nightside emission, and better constrain the nightside cloud thickness than our Spitzer data could. On the other hand, we found that our simulated MIRI LRS observations could not make any better constraints on the allowed variation in thickness than the Spitzer data. A spectrally resolved analysis, as opposed to our band-integrated approach here, may prove more effective though.

\section*{Acknowledgements}
We would like to thank Dr. N\'{e}stor Espinoza for sharing techniques which aided our analysis. We would also like to thank our anonymous referee for encouraging comments and suggestions that improved this work.

This work is based on observations made with the \textit{Spitzer Space Telescope}, which was operated by the Jet Propulsion Laboratory, California Institute of Technology under contract with NASA. This research has made use of the NASA Exoplanet Archive, which is operated by the California Institute of Technology, under contract with the National Aeronautics and Space Administration under the Exoplanet Exploration Program. We also made use of, and appreciate the contributors to NASA's Astrophysics Data System and the Spitzer Heritage Archive.

\facilities{Exoplanet Archive, Spitzer (IRAC)}

\software{\texttt{astropy} \citep{astropy2013, astropy2018}, 
\texttt{BATMAN} \citep{Kreidberg15}, 
\texttt{emcee} \citep{Mackey2013}, 
\texttt{Matplotlib} \citep{Hunter2007}, 
\texttt{Numpy} \citep{Harris2020}, 
and the Python programming language. } 

\appendix 

\begin{figure*}[h!]
    \centering
    \includegraphics[width=\textwidth]{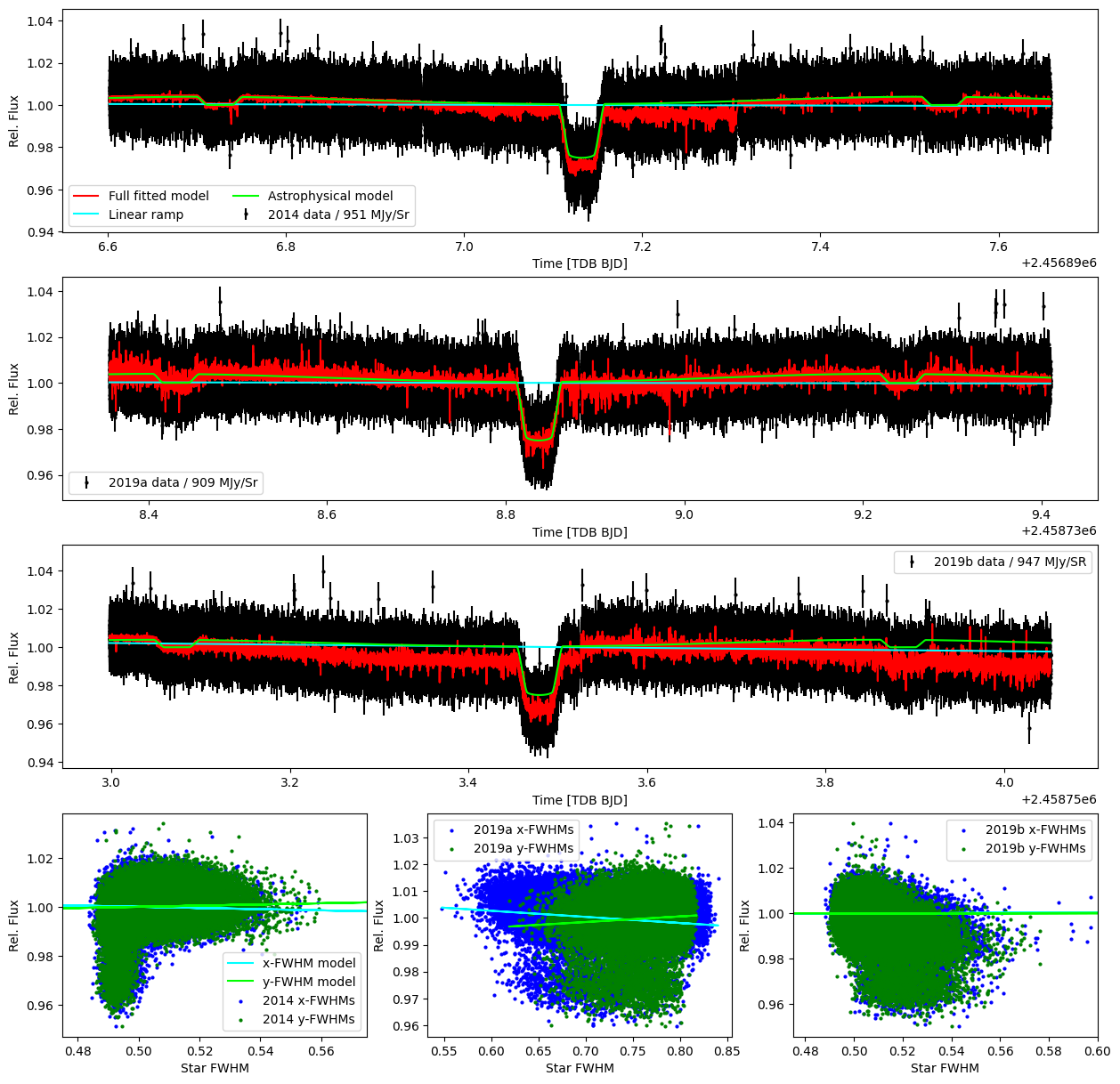}
    \caption{Plots of our full data sets along with each fitted model. The top three rows show the normalized, unbinned light curve of the 2014, 2019a, and 2019b visits, respectively. The normalization factor used for each is given in their corresponding legend. In these top three panels, the red line represents the full best-fit model including astrophysical and systematic components, the cyan line represents the best-fit linear ramp, and the lime line represents the best-fit astrophysical model. The bottom row plots the normalized, unbinned flux against the x- and y-FWHMs of WASP-43 on our images for each visit. The best-fit x-FWHM model is shown in cyan, and the best fit y-FWHM model is shown in lime, for each visit. These best-fit models are all from our fully shared fit case, which provided the best fit.}
    \label{fig:rawplots}
\end{figure*}

\begin{figure*}[h!]
    \centering
    \includegraphics[width=\textwidth]{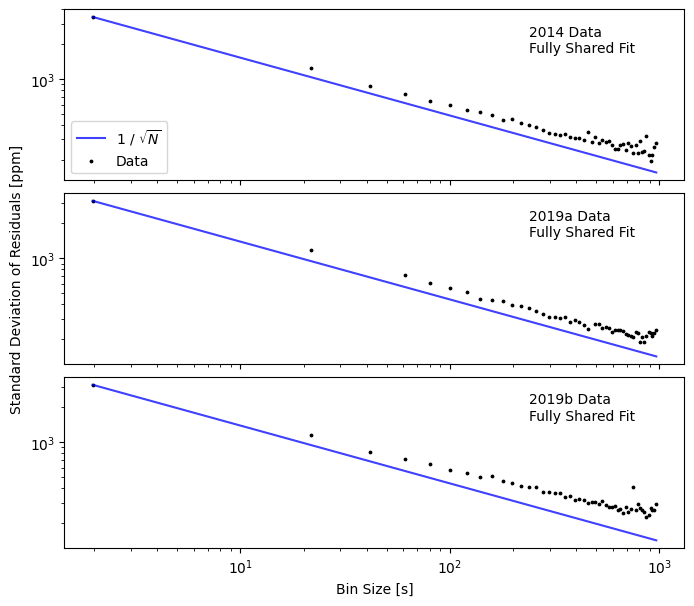}
    \caption{Plot of the standard deviations of the residuals as a function of bin size for each visit, after applying all components of our fully shared fit model. For each data set, the standard deviations decrease roughly as 1 / $\sqrt{N}$ as the bin size is increased, suggesting there is not much excess red noise left over in the detrended data, and each visit's plot looks nearly the same.
    }
    \label{fig:rednoisechecks}
\end{figure*}

Figure~\ref{fig:rawplots} shows our each of our full light curves, with the normalization value given in the caption, and the corresponding best fit astrophysical and systematic models for that visit. Figure~\ref{fig:rednoisechecks} shows “Allan variance” plots for each visit, showing how the residual scatter from our best fit models compares to pure white noise expectations. 

\newpage

\bibliographystyle{aasjournal}
\bibliography{journal}

\end{document}